\documentclass{ws-mpla}
\usepackage{undertilde}
\usepackage{multirow}
\usepackage{slashed}

\begin{document}

\markboth{Benjamin Fuks}
{Beyond the Minimal Supersymmetric Standard Model: from theory to phenomenology}

\catchline{}{}{}{}{}

\title{BEYOND THE MINIMAL SUPERSYMMETRIC STANDARD MODEL: FROM THEORY TO
PHENOMENOLOGY}

\author{\footnotesize BENJAMIN FUKS}

\address{Institut Pluridisciplinaire Hubert Curien/D\'epartement 
    Recherches Subatomiques, Universit\'e de Strasbourg/CNRS-IN2P3, 
    23 rue du Loess, F-67037 Strasbourg, France\\
    benjamin.fuks@iphc.cnrs.fr}

\maketitle


\begin{abstract}
  Thanks to the latest development in the field of Monte Carlo event generators and
  satellite programs allowing for a straightforward  implementation of
  any beyond the Standard Model theory in those tools, studying the property of
  any softly-broken supersymmetric theory is become an easy task. We illustrate
  this statement in the context of two non-minimal supersymmetric theories,
  namely the Minimal Supersymmetric Standard Model with $R$-parity violation and
  the Minimal $R$-symmetric Supersymmetric Standard Model and choose to probe
  interaction vertices involving a non-standard color structure and the sector
  of the top quark. We show how to 
  efficiently implement these theories in the {\sc Mathematica} package 
  {\sc FeynRules}  and use its interfaces to Monte Carlo tools for 
  phenomenological studies. For the latter, we employ the latest version of the
  {\sc MadGraph} program.

\end{abstract}

\ccode{11.30.Pb; 12.60.Jv; 13.85.Hd; 13.85.-t; 14.80.Ly.}

\newcommand{\alpgen}{{\sc Alpgen}}
\newcommand{\comix}{{\sc Comix}}
\newcommand{\calchep}{{\sc CalcHep}}
\newcommand{\comphep}{{\sc CompHep}}
\newcommand{\helac}{{\sc Helac}}
\newcommand{\mgme}{{\sc MadGraph/MadEvent}}
\newcommand{\sherpa}{{\sc Sherpa}}
\newcommand{\whizard}{{\sc Whizard}}
\newcommand{\lanhep}{{\sc LanHep}}
\newcommand{\feynrules}{{\sc FeynRules}}
\newcommand{\sarah}{{\sc Sarah}}
\newcommand{\ie}{{\textit{i.e.}}}
\newcommand{\python}{{\sc Python}}
\newcommand{\feynarts}{{\sc FeynArts}}
\newcommand{\formcalc}{{\sc FormCalc}}
\newcommand{\gosam}{{\sc GoSam}}
\newcommand{\madgraph}{{\sc MadGraph}}
\newcommand{\eg}{{\textit{e.g.}}}
\newcommand{\aloha}{{\sc Aloha}}
\newcommand{\helas}{{\sc Helas}}
\newcommand{\be}{\begin{equation}} 
\newcommand{\ee}{\end{equation}} 
\def\bsp#1\esp{\begin{split}#1\end{split}} 
\newcommand{\thetabar}{{\bar\theta}} 
\renewcommand{\d}{{\rm d}} 
\newcommand{\alphadot}{{\dot\alpha}} 
\newcommand{\Dbar}{{\overline D}} 
\newcommand{\del}{{\partial}} 
\newcommand{\mathematica}{{\sc Mathematica}}
\def\bpm{\begin{pmatrix}}
\def\epm{\end{pmatrix}}
\newcommand{\lag}{{\cal L}}
\newcommand{\hc}{{\rm h.c.}}
\newcommand{\herwig}{{\sc Herwig}}
\newcommand{\pythia}{{\sc Pythia}}
\def\lpp{{\lambda^{\prime\prime}}}
\newcommand{\spheno}{{\sc SPheno}}
\newcommand{\delphes}{{\sc Delphes}}
\newcommand{\madanalysis}{{\sc MadAnalysis}}

\begin{flushright}
  \vspace{-12.25cm} IPHC-PHENO-12-01\vspace{11.25cm} 
\end{flushright}
\section{Introduction}	
Many extensions to the Standard Model of particle physics have been proposed
over the last forty years. In particular, weak scale supersymmetry, and
especially its minimal version dubbed the Minimal Supersymmetric Standard Model
(MSSM) \cite{Nilles:1983ge,Haber:1984rc}, is one of the most theoretically and
experimentally studied alternatives of the Standard Model. Supersymmetry is a
symmetry relating particles with opposite statistics. In the MSSM, one single 
fermionic (bosonic) superpartner is assigned to each of the Standard
Model bosonic (fermionic) degree of freedom, with the exception of the Higgs
sector which contains two doublets of scalar Higgs fields, together with their
fermionic superpartners. In this way, several conceptual
problems of the Standard Model are addressed, such as the large hierarchy
between the Planck and the electroweak scale, gauge coupling unification at high
energy or the existence of a candidate for dark matter. However, since the
superpartners of the Standard Model particles have not yet been observed,
supersymmetry has to be broken at low energies. In addition, in order to remain
a viable solution to the hierarchy problem, this breaking has to be soft,
yielding supersymmetric masses around the TeV scale. Therefore, the quest for
supersymmetric particles is one of the main topics of the current experimental
program at the Large Hadron Collider (LHC) at CERN, which is currently exploring
the electroweak scale.

The general purpose LHC experiments, ATLAS and CMS, are currently
pushing the limits on the masses of the supersymmetric particles to a higher and
higher range \cite{ATLAS,CMS}. However, most of the analyses only hold in the
context of the so-called constrained MSSM, where $R$-parity is conserved and
where the 105 additional free parameters of the most general ($R$-parity
conserving) version of the MSSM are reduced to a set of four parameters and a
sign, assuming an organizing principle based on unification at high energy. In
contrast, there exists a vast variety of (non-minimal) supersymmetric models
which can be valuable to be investigated both from the point of view of the
theorist and the one of the experimentalist.

Studying the hadron-collider phenomenology of these non-minimal models
requires the use of Monte Carlo event generators in order to describe both 
backgrounds and new physics signals. These tools efficiently match a
proper modeling of the strong interactions (parton showering, fragmentation and
hadronization) with the calculation of matrix elements underlying the considered
hard-scattering processes where new physics is expected to appear. For this
reason, activities in the field of Monte Carlo simulations have been rather
intense over the last twenty years, resulting in many advances 
through the release of automated tree-level matrix-element generators, such as
\alpgen\ \cite{Mangano:2002ea}, \comix\ \cite{Gleisberg:2008fv},
\comphep/\calchep\ \cite{Pukhov:1999gg,Boos:2004kh,Pukhov:2004ca}, \helac\
\cite{Cafarella:2007pc}, \mgme\
\cite{Stelzer:1994ta,Maltoni:2002qb,Alwall:2007st,Alwall:2008pm,Alwall:2011uj},
\sherpa\ \cite{Gleisberg:2003xi,Gleisberg:2008ta} and \whizard\
\cite{Moretti:2001zz,Kilian:2007gr}.

Computing predictions for a given new physics theory with the help of the
above-mentioned tools requires the implementation of the complete set of
associated Feynman rules in these programs.
This task has been alleviated by packages such 
as \lanhep\ \cite{Semenov:1998eb,Semenov:2008jy},
\feynrules\
\cite{Christensen:2008py,Christensen:2009jx,Christensen:2010wz,Duhr:2011se} or
\sarah\ \cite{Staub:2009bi,Staub:2010jh} which start from the Lagrangian of the
theory and export the associated Feynman rules to one or several event
generators in an automated fashion. 
However, whereas all these tools can address the implementation of the
Lagrangian in the usual space-time, only the \feynrules\ package is suitable to
perform \textit{ab initio} computations within the superspace formalism
\cite{Salam:1974yz,Ferrara:1974ac}, a natural framework for supersymmetric model
building.

In a softly broken supersymmetric theory, the field content is embedded into
chiral and vector supermultiplets for the matter and gauge sector, respectively,
and the Lagrangian is written as a sum of four terms. The first two consist in
kinetic terms for the chiral and vector supermultiplets of the
model. Assuming renormalizability, these terms are entirely fixed
by supersymmetry and gauge invariance and are hence model-independent. The third
piece of the Lagrangian describes the interactions among the chiral
supermultiplets and is driven by the so-called superpotential, whilst the last piece
contains the supersymmetry-breaking terms. These two last parts of the
Lagrangian are both model-dependent. In the superspace formalism, the Lagrangian
is written in terms of superfields, in contrast to the component field formalism
where it is expressed in terms of the usual scalar, fermionic and vector fields
of particle physics. The main advantage is related to simpler and very compact
superfield expressions, in contrast to their component field counterparts. This
renders the implementation of softly broken supersymmetric Lagrangians 
in high-energy physics tools much more straightforward and bypasses many
possible sources of mistakes.

The superspace module of \feynrules\ contains a set of
functions generating automatically the kinetic terms of the Lagrangian
(expressed in terms of superfields).
Subsequently, the implementation of any supersymmetric theory in the \feynrules\
package, and therefore in any of the Monte Carlo event generators interfaced to
it, consists only in the setting of the superfield content, the model parameters
and the gauge symmetries of the theory, together with providing the model
dependent parts of the Lagrangian, \ie, the supersymmetry-breaking Lagrangian and the
superpotential (the latter being expressed in terms of superfields). Once
implemented, the 
Lagrangian is automatically expanded in terms of component fields by \feynrules,
these objects being the ones handled by the Monte Carlo event generators. The
Feynman rules are then extracted and the vertices are passed to the different
interfaced tools. 

This procedure allows for phenomenological investigations
of a large class of supersymmetric models whose full implementation in a Monte Carlo
event generators was considered too involved so far. Currently, dedicated
interfaces exist to the \comphep/\calchep, \feynarts/\formcalc\
\cite{Hahn:1998yk,Hahn:2000kx}, \mgme, \sherpa\ and \whizard\ programs. In
addition, a Universal \feynrules\ Output (UFO)
\cite{Degrande:2011ua} of the model can be created. In this case, all the model
information is exported as a set of \python\ classes and objects representing
particles, parameters and vertices. The produced \python\ library contains all
the interactions included in the Lagrangian, without any restriction on the
allowed Lorentz and/or color structures, in contrast to the other interfaces
which reject vertices not compliant with the (hard-coded) structures supported
by the corresponding programs. Presently, the UFO format is used by the \madgraph\
(version 5) and the \gosam\ \cite{Cullen:2011ac} generators\footnote{Even if the
\gosam\ program is also
using the UFO as its standard model format, this tool is dedicated to
next-to-leading order computations, which goes beyond the scope of this work.}.

In this work, we present the phenomenological study of collider signatures
associated to interaction vertices with exotic color structures. We start by
describing how to implement into \feynrules\ two examples of supersymmetric
theories beyond the constrained MSSM in the most automated possible way, to
extract the UFO libraries and to use it then with the \madgraph\ Monte Carlo
event generator for phenomenological investigations of signatures specific to
each model. The choice of the \madgraph\ generator is driven by the color
structures which we desire to probe, since among the above-mentioned tools, only 
\madgraph\ 5 is capable of handling non-standard color or Lorentz structures,
using, among others, the flexibility of the UFO model format,
which is handled internally by \aloha\ \cite{deAquino:2011ub}, an application
automatically writing \helas\ \cite{Murayama:1992gi,Hagiwara:2008jb,%
Hagiwara:2010pi,Mawatari:2011jy} libraries
from Feynman rules corresponding to the model under consideration.

In our first example, we extend the constrained ($R$-parity conserving) MSSM
by allowing superpotential and
supersymmetry-breaking Lagrangian terms in which $R$-parity is broken
\cite{Barbier:2004ez}. We choose to focus on a set of particular vertices
included in that model which couple three particles lying in the fundamental
representation of $SU(3)_c$ through a totally antisymmetric tensor of
three color (anti)triplet indices. For our second example, 
we turn to phenomenological investigations of scalar particles lying in the
adjoint representation of the $SU(3)_c$ gauge group, as predicted by  
hybrid supersymmetric theories containing representations of both the $N=1$
and $N=2$ superalgebras \cite{Fayet:1975yi,AlvarezGaume:1996mv,%
Choi:2008pi,Choi:2008ub,Choi:2009jc,Choi:2010gc,Schumann:2011ji} or by
$R$-symmetric
supersymmetric models \cite{Salam:1974xa,Fayet:1974pd,Kribs:2007ac,Choi:2010an}.
In those models, the $SU(3)_c$ vector supermultiplet of the MSSM is supplemented
by a chiral supermultiplet, forming hence a complete gauge supermultiplet of a $N=2$
supersymmetry lying in the adjoint representation of the $SU(3)_c$ gauge group.
This latter supermultiplet contains a vector field (the gluon), two
(two-component) fermions mixing to a Dirac gluino field (in contrast to
the Majorana counterpart included in the MSSM) and one scalar particle, commonly
dubbed as the sgluon field. Even if pairs of sgluons interact with one or two
gluons through the usual covariant derivative of $SU(3)_c$, hybrid and
$R$-symmetric supersymmetric theories also
predict a coupling between a single sgluon and a pair of gluons through
a dimension-five operator generated by a loop
of squarks, the color part of the interaction vertex being the
symmetric tensor of $SU(3)$ \cite{Plehn:2008ae}. Since the QCD sector
is the same in both the $R$-symmetric and hybrid supersymmetric models, we
choose to focus, in the following, only on the $R$-symmetric theories for the
sake of the example.

Even if we decide to focus on non-minimal softly-broken supersymmetric 
theories predicting interaction vertices with
peculiar color structures, a broad class of extensions of the MSSM contains
additional non-renormalizable interactions in the K\"ahler potential and in the
superpotential, leading to non-standard Lorentz structures.
These new interactions appear, \eg\, in the Higgs sector
\cite{Brignole:2003cm,Carena:2009gx,Antoniadis:2009rn}, when we consider
possible new physics effects through higher-dimensional operators. These models
can be directly implemented into the superspace module of \feynrules\ and
consequently exported to Monte Carlo tools for further phenomenological
investigations in the same way as for the two examples considered in this work.
In this case, Monte Carlo programs containing a hard-coded set of supported
Lorentz structures are again not suitable if the relevant structures
are too exotic. Once again, the \madgraph\ 5 program, using the UFO and
\aloha, allows for the implementation of any (renormalizable or not)
interaction vertex regardless of the complexity of the Lorentz structure.

This paper is organized as follows. In Section \ref{sec:feynrules}, we recall
the general features of softly-broken supersymmetric Lagrangians and describe
the two considered non-minimal supersymmetric theories, namely the MSSM with
$R$-parity violation and the minimal $R$-symmetric supersymmetric model, and
their implementation within the superspace module of
\feynrules\ starting from the existing
MSSM implementation. In Section \ref{sec:UFOMG}, we present the extraction of
the UFO model format and its linking to \madgraph\ 5, whilst Section
\ref{sec:bench} focuses on the adopted benchmark scenarios and the description
of the high energy physics process of interest. Section
\ref{sec:pheno} is dedicated to our phenomenological analyses and our
conclusions are finally given in Section \ref{sec:conclusions}.

\section{Softly-broken supersymmetric theories and their implementation in
\feynrules} \label{sec:feynrules}

In this Section, we describe the main features of softly-broken
supersymmetric theories and how to implement the corresponding Lagrangians into
the superspace module of \feynrules. For further information, we refer to the
manual of \feynrules\ \cite{Christensen:2008py} and to the details of its
superspace module \cite{Duhr:2011se}. We then describe the two 
non-minimal supersymmetric models which we propose to investigate,
\ie, the Minimal Supersymmetric Model with
$R$-parity violation and the Minimal $R$-symmetric Supersymmetric Standard
Model, and finally address their implementation within \feynrules.

\subsection{A generic supersymmetric theory in the superspace formalism}
\label{sec:superspace}

\subsubsection{Superfields}

Supersymmetric theories are more elegantly formulated within the superspace
formalism \cite{Salam:1974yz,Ferrara:1974ac}. Superspace is an extension of the
ordinary space-time where we supplement the usual space-time coordinates
$x^\mu$ by the variables $\theta$ and $\thetabar$, 
transforming as two-component Weyl fermions
with opposite chirality (forming thus a Majorana spinor
$(\theta^\alpha,\thetabar_\alphadot)$).
A superfield $\Phi$ is any function of the superspace coordinates and can
generally expanded as a (finite) Taylor series with respect to the variables
$\theta$ and $\thetabar$,
\be\bsp
  \Phi(x,\theta,\bar\theta) = &\ z(x) 
    + \theta \!\cdot\! \xi(x) + \bar \theta \!\cdot\! \bar \zeta(x) 
    + \theta \!\cdot\! \theta f(x) + \bar \theta \!\cdot\! \bar \theta g(x)   
    + \theta \sigma^\mu \bar \theta v_\mu(x)
\\ &\quad 
    + \bar \theta \!\cdot\! \bar \theta\ \theta \!\cdot\! \omega(x) 
    + \theta \!\cdot\! \theta\ \bar \theta \!\cdot\! \bar \rho(x) 
    + \theta \!\cdot\! \theta\ \bar \theta \!\cdot\! \bar \theta d(x) \ . 
\esp \label{eq:genSF} \ee
The various coefficients of this expansion are the component fields of the
superfield $\Phi$ and form a so-called supermultiplet. The fields $z$, $f$, $g$
and $d$ are thus complex scalar fields,
the spinors $\xi$, $\zeta$, $\omega$ and $\rho$ are complex Weyl fermions and
$v_\mu$ is a complex vector field. In the expression above, we have introduced
the four-vector built from the Pauli matrices $\sigma^\mu$. All the conventions,
and in particular those related to spinors (invariant products, position of the
indices, $\ldots$), strictly follow those presented in Ref.\
\refcite{Duhr:2011se} and Ref.\ \refcite{FuksRausch}.

The unconstrained superfield $\Phi$ of Eq.\ \eqref{eq:genSF} provides a
reducible representation of the supersymmetric algebra, the number of degrees of
freedom embedded in the expansion of $\Phi$ being 
too large with respect to those included in the supermultiplets representing
the $N=1$ supersymmetric algebra. Consequently, constraints are imposed on the
generic superfield $\Phi$ so that the number of independent component fields is
drastically reduced. The two types of superfields necessary to construct most of
the phenomenologically relevant supersymmetric theories are chiral superfields,
which include one complex scalar and one two-component fermionic component, and
vector superfields whose component fields are one real massless vector field
and the associated Majorana fermion.

Left and right-handed chiral superfields $\Phi_L$ and $\Phi_R$ satisfy the
constraints
\be
  \Dbar_\alphadot \Phi_L (x,\theta,\bar\theta) = 0 
  \quad \text{and} \quad
  D_\alpha \Phi_R (x,\theta,\bar\theta) = 0 \ ,
\label{eq:chiralconstraint}\ee 
where the superderivatives $D_\alpha$ and $\Dbar_\alphadot$ read, in our
conventions \cite{FuksRausch}, 
\be \bsp
  D_\alpha = \frac{\del}{\del\theta^\alpha} - i \sigma^\mu{}_{\alpha\alphadot}
    \bar \theta^\alphadot \del_\mu 
  \quad\text{and}\quad
  \Dbar_\alphadot =  \frac{\del}{\del \bar\theta^\alphadot} - i\theta^\alpha
    \sigma^\mu{}_{\alpha\alphadot}\del_\mu \ .
\esp\label{eq:superD} \ee
Under the constraints of Eq.\ \eqref{eq:chiralconstraint}, the number of degrees of
freedom included in the expansion of the superfields $\Phi_L$ and $\Phi_R$ with
respect to the variables $\theta$ and $\thetabar$ match those of the matter
supermultiplets. The expansions of these superfields are indeed given by
\be\bsp
  \Phi_L(y,\theta) = &\ \phi(y) + \sqrt{2} \theta \!\cdot\! \psi(y) - \theta
    \!\cdot\! \theta F(y) \ , \\
  \Phi_R(y^\dagger,\bar\theta) = &\ \phi(y^\dagger) + \sqrt{2} \bar\theta
    \!\cdot\! \bar\psi(y^\dagger) - \bar\theta \!\cdot\! \bar\theta F(y^\dagger) \ .
\esp\ee
In these two equations, we have introduced the reduced variable 
$y^\mu = x^\mu - i\theta
\sigma^\mu \thetabar$. As stated above, the expansion of the superfields
$\Phi_L$ and $\Phi_R$ includes hence a complex scalar field $\phi$, a Weyl 
fermion $\psi$ and an auxiliary additional scalar field $F$, which is
necessary to restore equality between the number of bosonic and
fermionic degrees of freedom for off-shell supermultiplets. 

Chiral superfields can be declared in \feynrules\ model files in a way
similar to the one employed for the declaration of ordinary fields. This extends
the original \feynarts\ conventions to the superfield level. Superfields with
the same quantum numbers are collected into classes and each class is
implemented as a list of \mathematica\footnote{
{\sc Mathematica} is a registered trademark of Wolfram Research Inc.}
replacement rules. For
example, a left-handed superfield $\Phi$ could be implemented as
\begin{verbatim}
  CSF[1] == {
    ClassName -> PHI,
    Chirality -> Left,
    Scalar    -> z
    Weyl      -> psi,
    Indices   -> {Index[Colour]} 
  }
\end{verbatim}
This declares simultaneously a left-handed chiral superfield \texttt{CSF[1]}
represented by the variable \texttt{PHI} and the associated right-handed
chiral superfield $\Phi^\dag$. The latter can then be used in \feynrules\
through the (automatically created) symbol \texttt{PHIbar}. 
In addition, the implemented superfield carries an index labeled by {\tt
Colour}, illustrating the fact that it lies in a non-trivial representation of
the QCD gauge group (see below). The scalar and fermionic component fields of
$\Phi$ are represented by the symbols \texttt{z} and \texttt{psi}, respectively,
and it is important to note that these two fields must be independently declared
in the \feynrules\ model file. In contrast, the auxiliary $F$-field is
internally handled and the user does not have to take care of it. For
instance, the scalar component field could be included in the model file as 
\begin{center}
\begin{verbatim}
  S[1] == { 
    ClassName     -> z,
    SelfConjugate -> False,
    Indices       -> {Index[Colour]}
  } 
\end{verbatim}
\end{center}
We emphasize that the options, such as {\tt Indices}, which are shared by the
component field class and the superfield class must be identical. A complete
description of the particle and superfield classes and properties can be found
in Ref.\ \refcite{Christensen:2008py} and Ref.\ \refcite{Duhr:2011se}.

Vector superfields describe the representations of the $N=1$ supersymmetric
algebra containing one massless gauge boson together with the corresponding
fermionic degree of freedom. These multiplets are embedded into vector
superfields defined by the reality condition
\be
  V = V^\dagger \ .
\ee
However, this last condition leads to numerous non-physical fields which have to be
integrated out. In a suitable gauge, the Wess-Zumino gauge, these unphysical 
components are eliminated already at the level of
the expansion of the superfield $V$ in terms of the variables $\theta$ and
$\thetabar$, 
\be
  V = \theta \sigma^\mu \bar \theta v_\mu + 
    i \theta \!\cdot\! \theta\ \thetabar \!\cdot\! \bar \lambda - 
    i \bar \theta \!\cdot\! \thetabar\ \theta \!\cdot\! \lambda + 
    \frac12 \theta \!\cdot\! \theta\ \thetabar \!\cdot\!\thetabar D \ .
\ee
This gauge choice is adopted in the rest of this paper.
The component fields included in the expansion of the superfield $V$ are a 
real vector field
$v_\mu$, a Majorana spinor $(\lambda_\alpha,\bar\lambda^\alphadot)$ and an
additional auxiliary real scalar field $D$ ensuring equal numbers of fermionic and
bosonic degrees of freedom when going off-shell.

This superfield can be
declared in \feynrules\ model files in a similar fashion as for the chiral
superfield. Taking the example of the QCD gauge group, one could have  
\begin{verbatim}
  VSF[1] == { 
    ClassName  -> GSF, 
    GaugeBoson -> G,
    Gaugino    -> gow,
    Indices    -> {Index[Gluon]}
  }  
\end{verbatim}
This list of \mathematica\ replacement rules
assigns the symbol \texttt{GSF} to the vector superfield \texttt{VSF[1]} 
and defines the related physical components fields by the variables
\texttt{G}, representing the gluon, and \texttt{gow}, corresponding to the 
two-component gluino spinor. As for the chiral superfield case, both the gluon
and gluino fields must be properly declared in the model file and the auxiliary
$D$-field is treated internally by \feynrules. As vector superfields and their 
component fields are
associated to the mediation of the gauge interactions, they naturally lie
in the adjoint representation of the gauge group. This is
emphasized by the presence of an index denoted \texttt{Gluon}\footnote{Strong
interactions play a special role in Monte Carlo programs related to the
simulation of hadron collisions. Consequently, 
the names of the quantities related to the QCD gauge group, such as 
the index related to its adjoint representation, follow the strict conventions
depicted in Ref.\ \refcite{Christensen:2008py}.} among the
properties of the superfield {\tt VSF[1]}, and this index is related to the
adjoint representation of $SU(3)_c$ once this group is defined. This last
task could be performed through the declaration 
\begin{verbatim}
  SU3C == 
  { 
    Abelian           -> False,
    Superfield        -> GSF,
    Representations   -> { {T,Colour}, {Tb,Colourb} },
    CouplingConstant  -> gs, 
    StructureConstant -> f,
    DTerm             -> dSUN
  } 
\end{verbatim}
This last \mathematica\ command declares a variable {\tt SU3C} representing the
$SU(3)_c$ gauge group and associated to the superfield {\tt GSF} through the 
{\tt Superfield} option of the gauge group class. This mapping
internally defines the indices carried by the superfield {\tt GSF} as those
related to the adjoint representation of the gauge group. In addition, two other
representations are declared, the symbols {\tt Colour} and {\tt Colourb} being 
the fundamental and anti-fundamental representation indices and {\tt T} and {\tt Tb} 
the corresponding representation matrices.
The declaration of the gauge group also assigns the symbol
{\tt gs} to the strong coupling constant and {\tt dSUN} and {\tt f} to the
symmetric tensor and antisymmetric structure constants of $SU(3)$, respectively. This
allows to use consistently build-in \feynrules\ functions computing
automatically, \eg, the superfield strength tensor or the chiral superfield
kinetic terms. 

\subsubsection{Lagrangians}\label{sec:theolag}
A generic supersymmetric model is defined by fixing the superfield content 
and the gauge symmetries of the theory. We consider a simple gauge group $G$ and 
denote by $g$ the related gauge coupling constant.  
We introduce a set of irreducible representations ${\cal R}$
of the corresponding Lie algebra ${\mathfrak g}$ which are spanned by the hermitian
matrices $T^a$, and to each matrix, associate a vector superfield $V^a$.
The matter sector consists in a set of matter supermultiplets
described by left-handed chiral superfields $\Phi^i$ lying in the
representation ${\cal R}^i$ of ${\mathfrak g}$, whilst the gauge sector is
described by a vector superfield $V=V^a T_a$.
In the case of a semi-simple gauge group, \ie, when the gauge group consists in
a direct product of several simple gauge groups, it
is sufficient to generalize the previous setup by associating one vector
superfield $V$ to each direct factor $G^j$ of the gauge group. The chiral
superfields of the theory then lie in a given representation of the algebras
corresponding to each of the $G^j$ groups.

In their most general (non-renormalizable) version, softly-broken supersymmetric
Lagrangians can be expressed entirely with the help of three fundamental 
functions of the superfields of the theory \cite{Cremmer:1978hn,Zumino:1979et,%
Cremmer:1982wb,Cremmer:1982en}, namely the K\"ahler potential $K$, the
superpotential $W$ and the gauge kinetic function $h_{ab}$, given together with
the soft supersymmetry breaking terms. The
K\"ahler potential is a real function of the chiral and anti-chiral superfield
content of the theory and contains kinetic terms describing its dynamics. The
superpotential is an holomorphic function of the chiral superfields and
describes their interactions not driven by the N\oe ther procedure leading to
the covariantization of the Lagrangian both with respect to gauge and
supersymmetric transformations. The gauge kinetic function, symmetric in the 
two adjoint gauge indices $a$ and $b$, is also an holomorphic
function of the chiral superfields, and addresses the kinetic terms related to the vector
supermultiplets of the theory. Let us note that the K\"ahler potential and
the gauge kinetic function have a very simple form if one assumes
renormalizability. In this case, the Lagrangian is much simpler and its
implementation in \feynrules\ can be automated (see below). We however first
turn to the most general case.

Putting everything together, (non-renormalizable) softly-broken
supersymmetric Lagrangian reads
\be \bsp 
  {\cal L} =&\  
    \int \d^2 \theta\ \d^2 \thetabar\ \frac12 \Big[ K(\Phi, \Phi^\dag 
      e^{-2gV}) + K(e^{-2gV} \Phi, \Phi^\dag) \Big] \\ 
  &\quad  
    + \frac{1}{16 g^2}\int \d^2 \theta\ h_{ab}(\Phi) W^{a\alpha} W^b_\alpha 
    + \frac{1}{16 g^2}\int \d^2 \thetabar\ h_{ab}^\star(\Phi^\dag) 
      \overline{W}^a_\alphadot \overline{W}^{b\alphadot}  \\
    &\quad + \int \d^2 \theta \ W(\Phi) 
       + \int \d^2 \thetabar \ W^\star(\Phi^\dag)  + 
    {\cal L}_{\rm soft}\ , 
\esp \label{eq:lsusy}\ee 
where an implicit sum over the chiral and gauge superfields $\Phi$ and $V$ is
understood, if relevant, 
and where $W^\star$ and $h^\star$ are the functions conjugate to $W$
and $h$, respectively. In this last equation, we have introduced the
soft-supersymmetry breaking Lagrangian ${\cal L}_{\rm soft}$ and the
spinorial superfield strength tensors. In the conventions of Ref.\
\cite{FuksRausch} they read,
\be
  W_\alpha= -\frac14 \Dbar \!\cdot\! \Dbar \ e^{2gV} D_\alpha e^{-2gV} 
  \quad  \text{and}\quad 
  \overline W_\alphadot = -\frac14  D \!\cdot\! D \ e^{-2gV} \Dbar_\alphadot
    e^{2gV} \ , 
\label{eq:superfieldstrength}\ee
where the superderivatives have been defined in Eq.\ \eqref{eq:superD}.

The general Lagrangian of Eq.\ \eqref{eq:lsusy} can be included in a
straightforward manner in \feynrules\ model files 
using the strengths of the superspace module. One possible implementation would
proceed in two steps. Firstly, the superpotential, expressed in terms of
superfields, is stored in the variable {\tt
SuperPot} and the supersymmetry-breaking Lagrangian, generally expressed in
terms of component fields\footnote{For a soft-supersymmetry breaking Lagrangian
defined in terms of superfields and integration over the superspace, we refer to
Section \ref{sec:mrssm}.}, is implemented in the variable {\tt LSoft}. Secondly,
for the kinetic part of the Lagrangian, we define the quantities
\be
  \Omega = \frac12 \Big[ K(\Phi, \Phi^\dag e^{-2gV}) + K(e^{-2gV} \Phi,
    \Phi^\dag) \Big] \quad\text{and}\quad 
  \rho =  h_{ab}(\Phi) W^{a\alpha} W^b_\alpha \ ,
\ee
and implement them (in terms of superfields) in the \feynrules\ model file
within the variables {\tt Omega} and {\tt rho}, respectively. An
implementation of the Lagrangian of Eq.\ \eqref{eq:lsusy} is finally given by
\begin{verbatim}
  Lag =  Theta2Thetabar2Component[ Omega ]  + 
    Theta2Component[ rho + SuperPot ] + 
    Thetabar2Component[ HC[rho] + HC[SuperPot] ] + 
    LSoft;
\end{verbatim}
In this set of commands, we have introduced the \feynrules\ functions
\texttt{Theta2Component}, \texttt{Thetabar2Component},
\texttt{Theta2Thetabar2Component} allowing to perform 
the two- and four-dimensional 
integrations over the Grassmann variables $\theta$ and
$\thetabar$, together with the function
{\tt HC} which transforms an expression into the hermitian conjugate
counterpart. Let us note that the
left-handed and right-handed superfield strength tensors appearing in the
definition of $\rho$ can be cast within the functions
\texttt{SuperfieldStrengthL} and \texttt{SuperfieldStrengthR}. We refer to Ref.\
\refcite{Duhr:2011se} for more details.

The Lagrangian density of Eq.\ \eqref{eq:lsusy} still depends
on the unphysical $F$ and $D$ degrees of freedom of the theory and is
expressed in terms of Weyl fermions. In contrast, all Monte
Carlo event generators which could be used to investigate the phenomenology of
any softly-broken supersymmetric model require the usage of four-component
Majorana and Dirac (and not two-component) fermions and physical fields. The
\feynrules\ package provides a couple of automated functions allowing to render
the Lagrangian compliant with the requirements of the Monte Carlo programs. 

Firstly, the
unphysical fields can be integrated out of the Lagrangian by inserting the
solution of their equations of motion. This can be performed by employing 
the functions {\tt SolveEqMotionD} and {\tt SolveEqMotionF} for the $D$- and
$F$-terms, respectively,
\begin{verbatim}
    Lag = SolveEqMotionD[ Lag ];
    Lag = SolveEqMotionF[ Lag ];
\end{verbatim}

Secondly, the {\tt WeylComponents} option of the four-component fermion particle
class allows to link a four-component fermion to its two-component counterparts.
Set up in this way, \feynrules\ replaces any occurrence of a product of
Weyl spinors
by the corresponding product of four-component spinors. This task is achieved 
through
the automated function {\tt WeylToDirac}. However, before performing these
transformations,
care must be taken with the gauge group representations in which the considered
fields lie. Indeed, supersymmetric field theories are built only with the help of
left-handed chiral superfields. Subsequently, the right-handed parts of the
four-component fermions are embedded into the corresponding charge-conjugate 
left-handed chiral superfield, lying thus in the complex conjugate
representation of the gauge group. Mapping Weyl fermions to Dirac and Majorana
fermions requires then a specific treatment. Taking the example of the QCD
gauge group implemented above, it is sufficient to render the indices {\tt
Colour} and {\tt Colourb} equal and use the properties of the representation
matrices $\bar T = - T^\ast = - T^t$ to get rid of any track of the
anti-fundamental
representation of $SU(3)_c$, $T$ being the fundamental and $-T^\ast$ the
anti-fundamental representation matrices, \ie, the complex conjugate
representation. Hence, a possible implementation would
be
\begin{verbatim}
  Colourb = Colour;
  Lag = Lag /. { Tb[a_,i_,j_] -> -T[a,j,i] };
  Lag = WeylToDirac[ Lag ];
\end{verbatim}

For renormalizable theories, the K\"ahler potential and the gauge
kinetic functions take a trivial form
\be
   K(\Phi,\Phi^\dag) = \Phi^\dag \Phi \quad\text{and}\quad
  h_{ab}(\Phi) = \delta_{ab} \ ,
\label{eq:renormsusy}\ee
and the Lagrangian of Eq.\ \eqref{eq:lsusy} simplifies. The
complete model is then described by the superpotential and the
soft-supersymmetry breaking terms alone, the $\Omega$ and $\rho$ variables being
fully fixed by Eq.\ \eqref{eq:renormsusy}. In this case, these quantities 
can be computed
automatically with the help of the \texttt{CSFKineticTerms} and
\texttt{VSFKineticTerms} functions of \feynrules, 
the only task left to the user being the
implementation of the superfield content of the theory, the superpotential and the
soft-supersymmetry breaking Lagrangian. This renders the implementation of any
softly-broken supersymmetric theory in \feynrules, and therefore in the Monte
Carlo programs interfaced to it, almost entirely automated. The two examples
investigated in this work follow that approach, the tasks related to programming
which are left to the user being therefore minimal.

\subsection{The Minimal Supersymmetric Standard Model with $R$-parity violation}
\label{sec:rpv}
\subsubsection{Theoretical framework}
The MSSM is the simplest supersymmetric model, resulting from a direct
supersymmetrization of the Standard Model \cite{Nilles:1983ge,Haber:1984rc}. In
the most general form allowed by gauge invariance and renormalizability, 
the superpotential includes dangerous baryon-number ($B$) and
lepton-number ($L$) violating interactions predicting, \eg, 
fast proton decays, atomic parity violation or large flavor-changing neutral
currents \cite{Barbier:2004ez}. To circumvent this issue, a discrete symmetry is
imposed, namely $R$-parity, which forbids all the problematic terms.

However, neither $B$ nor $L$ conservations are fundamental \cite{'tHooft:1976up}
and there is no strong theoretical motivation to forbid the $R$-parity
violating terms of the MSSM superpotential. 
One could rather assume their existence and derive strong experimental
constraints on the corresponding coupling strengths. In general, among all the
possible $R$-parity violating terms, data tends to show a single coupling
dominance effect, \ie, only one single $R$-parity coupling can be
non-negligible at a time \cite{Dimopoulos:1988jw,Barger:1989rk}.

The MSSM is based on the same semi-simple gauge group as the Standard Model,
$G_{MSSM} \equiv SU(3)_c \times SU(2)_L \times U(1)_Y$. We associate one
vector superfield
to each of the direct factors of the gauge group. These vector superfields,
labeled by $V_G$, $V_W$ and $V_Y$, lie in
the adjoint representation of the corresponding simple Lie group and are singlets
under all the other factors of $G_{MSSM}$. They are given, together with their
representations under $G_{MSSM}$, by 
\be \bsp
  SU(3)_c \Leftrightarrow  &\ V_G = ({\utilde{\bf 8}}, {\utilde{\bf 1}}, 0)\
    ,\\
  SU(2)_L \Leftrightarrow  &\ V_W = ({\utilde{\bf 1}}, {\utilde{\bf 3}}, 0)\
    ,\\
   U(1)_Y \Leftrightarrow  &\ V_B = ({\utilde{\bf 1}}, {\utilde{\bf 1}}, 0)\
    .
\esp\label{eq:VSFmssm}\ee
The set of physical component fields includes, in addition to the Standard Model
gauge bosons, the gaugino partners, as shown in Table \ref{tab:gauge}.

\renewcommand{\arraystretch}{1.4}
\begin{table}[h]
  \tbl{The superfields included MSSM gauge sector. The gauge boson and gaugino
    components are given together with the representations under the
    $SU(3)_c \times SU(2)_L \times  U(1)_Y$ gauge group.}
  {\begin{tabular}{|c||c|c|c|c|}
    \hline
    Superfield & Gauge boson & Gaugino & Representation \\ \hline 
    \hline 
    $V_B$ & $B_\mu$ & $\widetilde B$ & $({\utilde{\bf 1}}, {\utilde{\bf 1}}, 0)$ \\
    $V_W$ & $W_\mu$ & $\widetilde W$ & $({\utilde{\bf 1}}, {\utilde{\bf 3}}, 0)$ \\
    $V_G$ & $g_\mu$ & $\widetilde g$ & $({\utilde{\bf 8}}, {\utilde{\bf 1}}, 0)$ \\
    \hline
  \end{tabular} \label{tab:gauge} }
\end{table}
\renewcommand{\arraystretch}{1}

The chiral sector of the theory consists of three generations of Standard
Model quarks and leptons embedded into three generations of chiral
supermultiplets, together with their squark and slepton partners,
\be\bsp
 &\ Q_L^i = ({\utilde {\bf 3}}, {\utilde{\bf 2}}, \frac16) \quad  ,\quad
    U_R^i = ({\utilde {\bf \bar 3}}, {\utilde{\bf 1}},-\frac23) \quad ,\quad
    D_R^i = ({\utilde {\bf \bar 3}}, {\utilde{\bf 1}}, \frac13) \\
 &\ L_L^i = ({\utilde {\bf 1}}, {\utilde{\bf 2}},-\frac12) \quad , \quad
    E_R^i = ({\utilde {\bf 1}}, {\utilde{\bf 1}}, 1) \quad  , \quad
    V_R^i = ({\utilde {\bf 1}}, {\utilde{\bf 1}}, 0) \ ,
\esp\label{eq:CSF1}\ee
where $i$ stands for a generation index and where we have indicated the
representations of the different superfields under the MSSM gauge
group. Let us note that since supersymmetric theories must be built only with
left-handed (and not right-handed) chiral superfields, the associated
superfields include the charge-conjugate right-handed Standard Model fermions
and lie thus in the corresponding complex-conjugate representations of
$G_{MSSM}$. For completeness, we have included the right-handed neutrino
superfields. However, they will be kept non-interacting with any
other superfield, even at the level of the superpotential.

In addition, (s)fermion superfields are supplemented with two Higgs chiral
supermultiplets, embedding thus 
two $SU(2)_L$ doublets of scalar Higgs fields accompanied by two
doublets of their fermionic Higgsino partners, 
\be
  H_D = ({\utilde{\bf 1}}, {\utilde{\bf 2}}, -\frac12)\quad ,\quad
  H_U = ({\utilde{\bf 1}}, {\utilde{\bf 2}},  \frac12) \ .
\label{eq:CSF2}\ee
This leads to electroweak symmetry breaking without
introducing chiral anomalies and to mass generation for both the up-type and
down-type fermions.
The component fields included in the full matter sector can be found in Table
\ref{tab:chiral}.

\renewcommand{\arraystretch}{1.1}
\begin{table}[h]
  \tbl{The superfields included in the MSSM chiral sector. The scalar and
two-component spinor components are given together with the
    representations under the $SU(3)_c \times SU(2)_L \times
    U(1)_Y$ gauge group. The superscript $c$ denotes charge conjugation.}
  {\begin{tabular}{|c||c|c|c|c|}
    \hline
    Superfield & Fermion & Scalar & Representation \\ \hline\hline 
    \multirow{4}{*}{$Q_L^i$}&&&\\
       &$q_L^i = \bpm u_L^i \\ d_L^i \epm$ & 
       $\tilde q^i_L = \bpm \tilde u^i_L\\  \tilde d^i_L\epm$ &
       $({\utilde {\bf 3}}, {\utilde{\bf 2}}, \frac16)$ \\&&&\\
    $U^i_R$ & 
       $u_R^{ic}$ & 
       $\tilde u_R^{i\dag}$ & 
       $({\utilde{\bf \bar 3}},{\utilde{\bf 1}}, -\frac23)$\\
    $D_R^i$ &
       $d_R^{ic}$ & 
       $\tilde d_R^{i\dag}$ & 
       $({\utilde{\bf \bar 3}},{\utilde{\bf 1}}, \frac13)$\\&&&\\
    \hline
    \multirow{4}{*}{$L_L^i$}&&&\\
       &$\ell_L^i = \bpm \nu_L^i \\ e_L^i\epm$ & 
       $\tilde \ell_L^i = \bpm\tilde \nu^i_L\\  \tilde e^i_L\epm$ &
       $({\utilde{\bf 1}}, {\utilde{\bf 2}}, -\frac12)$\\&&&\\
    $E_R^i$&
       $e_R^{ic}$ & 
       $\tilde e_R^{i\dag}$ & 
       $({\utilde{\bf 1}},{\utilde{\bf 1}}, 1)$\\&&&\\
    $V_R^i$&
       $\nu_R^{ic}$ & 
       $\tilde \nu_R^{i\dag}$ & 
       $({\utilde{\bf 1}},{\utilde{\bf 1}}, 0)$\\&&&\\
    \hline 
    \multirow{4}{*}{$H_D$}&&&\\
       &$\widetilde H_d = \bpm \widetilde H_d^0 \\ \widetilde H_d^- \epm$&
         $H_d = \bpm H_d^0 \\ H_d^- \epm$&
         $({\utilde{\bf 1}}, {\utilde{\bf 2}}, -\frac12)$ \\&&&\\
    \multirow{4}{*}{$H_U$}&&&\\
       &$\widetilde H_u = \bpm \widetilde H_u^+ \\ \widetilde H_u^0\epm$&
       $H_u = \bpm H_u^+ \\ H_u^0 \epm$ &         
        $({\utilde{\bf 1}}, {\utilde{\bf 2}}, \frac12)$ \\&&&\\
   \hline
   \end{tabular}\label{tab:chiral}}
\end{table}
\renewcommand{\arraystretch}{1}

As stated in Section \ref{sec:superspace}, the kinetic and gauge interaction
terms of the chiral and vector superfields are entirely fixed by gauge
invariance and supersymmetry. Considering the simplest model with trivial 
K\"ahler potential and gauge kinetic function (see Eq.\ \eqref{eq:renormsusy}),
the kinetic Lagrangian reads,
\be\bsp
  \lag = &\ \sum_{k = SU(3)_c, SU(2)_L,U(1)_Y}\Bigg(\int\d^2\theta 
        \frac{W_k^\alpha W^k_{\alpha}}{16 g^2} + \int\d^2\thetabar
       \frac{\overline W_{k\alphadot} \overline W^k_\alphadot}{16 g^2}
       \Bigg)
\\ &\
  + \sum_{\Phi=\text{chiral content}} \int \d^2\theta\d^2\thetabar
     \Phi^\dag \Big(e^{- Y g^\prime V_B} e^{-2 g_w \tilde V_W} e^{-2 g_s \tilde
      V_G}\Big) \Phi  \ ,
\esp\label{eq:lrpv}\ee
where the superfield strength tensors have been defined in Eq.\
\eqref{eq:superfieldstrength}. We have introduced the non-Abelian vector
superfields $\tilde V_W = V_{W^k} T_k$, $\tilde V_G = V_{G^a} T_a$, where the $SU(2)_L$
and $SU(3)_c$ generators $T_k$ and $T_a$ are taken, for each term in the
sum, in the representation appropriate to the considered chiral superfield
$\Phi$.
The three gauge coupling constants are denoted by $g^\prime$, $g_w$ and $g_s$. 

Non-gauge interactions among the chiral superfields introduced in Eq.\
\eqref{eq:CSF1} and Eq.\ \eqref{eq:CSF2} are driven by the
superpotential. The superpotential of our model is taken as the one of the
$R$-parity conserving
MSSM, denoted by $W_{RPC}$, supplemented by the so-called $\lambda$
terms\footnote{We do not consider in this work the bilinear Higgs-lepton
superfield mixing terms $\kappa_i L^i\!\cdot\! H_U$.},
\be \bsp
  W(\Phi) = &\ W_{RPC} \\ &\
    + \frac12 \lambda_{ijk} L_L^i \!\cdot\! L_L^j E_R^k
    + \lambda^\prime_{ijk} L_L^i \!\cdot\! Q_L^{j\ell} D^k_{R\ell}
    + \frac12 \lambda^{\prime\prime}_{ijk} \epsilon^{\ell m n} U_{R\ell}^i
        D_{Rm}^j D_{Rn}^k \ . 
\esp\label{eq:superwrpv}\ee
In this expression, we have explicitly indicated flavor indices ($i$, $j$ and
$k$), (anti)fundamental color indices ($\ell$, $m$, $n$) and the $SU(2)$
invariant product (as a `$\cdot$'). 

Similarly, additional $R$-parity violating trilinear scalar interactions derived
from the form of the superpotential can be included in the soft-supersymmetry
breaking Lagrangian, 
\be
  L_{\rm soft} = L_{{\rm s}, RPC} 
    - \frac12 T_{ijk} \tilde l_L^i \!\cdot\! \tilde l_L^j \tilde e_R^k
    - T^\prime_{ijk} \tilde l_L^i \!\cdot\! \tilde q_L^{j\ell} \tilde
       d^k_{R\ell}
    - \frac12 T^{\prime\prime}_{ijk} \epsilon^{\ell m n} \tilde u_{R\ell}^i
        \tilde d_{Rm}^j \tilde d_{Rn}^k + \hc  \ , 
\label{eq:lsoftrpv}\ee
where the coupling strengths are included in the $T$, $T'$ and $T''$ parameters
and the Lagrangian $L_{{\rm s}, RPC}$ is the $R$-parity conserving MSSM
soft-supersymmetry breaking Lagrangian.

\subsubsection{Implementation in \feynrules}

In the $R$-parity violating scenario described above, 
the superfield content of the theory is identical to
the one of the usual $R$-parity conserving MSSM. Moreover, all particle
mixings occurring after electroweak symmetry breaking 
are also left unchanged, since the additional terms
do not generate further mixing. Hence, the relations linking the gauge- and the
mass-eigenstate bases are left unchanged with respect to the $R$-parity 
conserving MSSM. Therefore, implementing the considered model into \feynrules\ 
can be performed 
very efficiently with a minimal effort. Implementing all the modifications 
in a file labeled {\tt rpv.fr}, it is then sufficient to load this file in
\feynrules\ simultaneously with the built-in MSSM model file, denoted by
{\tt mssm.fr},
\begin{verbatim}
  LoadModel["mssm.fr", "rpv.fr"];
\end{verbatim}
For details on the {\tt mssm.fr} file, which can be downloaded from the
\feynrules\ online model database, we refer to Ref.\
\refcite{Duhr:2011se}. 

The file {\tt rpv.fr} includes, on the one hand, 
the definition of the $R$-parity violating parameters $\lambda$, $\lambda'$, 
$\lpp$, $T$, $T'$ and $T''$. This follows
the standard rules for implementing instances of the particle class (see
the \feynrules\ manual). On the other hand, this file contains the
implementation of the model Lagrangian. The kinetic terms presented in 
Eq.\ \eqref{eq:lrpv} are directly implemented using the automated functions
{\tt CSFKineticTerms} and {\tt VSFKineticTerms},
\begin{verbatim}
  LagKin = Theta2Thetabar2Component[ CSFKineticTerms[ ] ] + 
    Theta2Component[ VSFKineticTerms[ ] ] +
    Thetabar2Component[ VSFKineticTerms[ ] ];
\end{verbatim}

The $R$-parity conserving MSSM superpotential is stored in the file {\tt
mssm.fr} in the variable {\tt SPot}. The complete Lagrangian containing the
interactions driven by the superpotential of Eq.\ \eqref{eq:superwrpv}
can then be implemented in the {\tt rpv.fr} file as  
\begin{verbatim}
  SupW = SPot + 
     LLLE[f1,f2,f3] Conjugate[PMNS[f4,f1]] *
         LL[1,f4] LL[2,f2] ER[f3] +
     LLQD[f4,f5,f3] Conjugate[CKM[f2,f5]] Conjugate[PMNS[f1,f4]] * 
         DR[f3,c1] (LL[1,f1] QL[2,f2,c1] - LL[2,f1] QL[1,f2,c1])
    1/2 LUDD[f1,f2,f3] Eps[c1,c2,c3] UR[f1, c1] DR[f2,c2] DR[f3,c3];
  LagW = Theta2Component[ SupW ] + Thetabar2Component[ HC[SupW] ];
\end{verbatim}
In the \mathematica\ commands above, the symbols {\tt LLLE}, {\tt LLQD} and
{\tt LUDD} stand for the $\lambda$, $\lambda'$ and $\lambda''$ parameters of the
superpotential, {\tt Eps} for
the fully antisymmetric tensor of rank three and {\tt LL}, {\tt ER}, {\tt QL},
{\tt UR} and {\tt DR} for the chiral superfields $L_L$, $E_R$, $Q_L$, $U_R$ and
$D_R$. For the first and second terms, the CKM and PMNS matrices
are explicitly included. This allows to compensate
further field redefinitions included in the model file (see Ref.\
\refcite{Duhr:2011se} for more information).

On the same footings, assuming that the soft-supersymmetry breaking Lagrangian
associated to the MSSM with $R$-parity conservation is implemented in the
variable {\tt LS}, (see {\tt mssm.fr}), the full supersymmetry-breaking
Lagrangian is implemented from the expression of Eq.\ \eqref{eq:lsoftrpv},
\begin{verbatim}
  Tsoft = TLLE[f1,f2,f3] Conjugate[PMNS[f4,f1]] * 
         LLs[1,f4] LLs[2,f2] ERs[f3] -
     TLQD[f4,f5,f3] Conjugate[CKM[f2,f5]] Conjugate[PMNS[f1,f4]] * 
         DRs[f3,c1] (LLs[1,f1] QLs[2,f2,c1]-LLs[2,f1] QLs[1,f2,c1]) - 
     1/2 TUDD[f1,f2,f3] Eps[c1,c2,c3] * 
         URs[f1, c1] DRs[f2,c2] DRs[f3,c3];
  LSoft = LS + Tsoft + HC[Tsoft]; 
\end{verbatim} 
where the symbols {\tt TLLE}, {\tt TLQD} and {\tt TUDD} stand for the $T$, $T'$
and $T''$ parameters of the soft-supersymmetry breaking Lagrangian and {\tt
LLS}, {\tt ERs}, {\tt QLs}, {\tt URs} and {\tt DRs} for the scalar component of
the $L_L$, $E_R$, $Q_L$, $U_R$ and $D_R$ superfields, respectively. Concerning
the presence of the CKM and PMNS matrices, we again refer to Ref.\
\refcite{Duhr:2011se}.

The complete model Lagrangian is thus given by
\begin{verbatim}
  Lag = LagKin + LagW + LSoft;
\end{verbatim}
In order to render this Lagrangian {\tt Lag} compliant with the requirements
of the Monte Carlo programs which are aimed to be used, the auxiliary $F$- and
$D$-fields must be integrated out. This can be performed with the help of the
{\tt SolveEqMotionD} and {\tt SolveEqMotionF} commands,
\begin{verbatim}
  Lag = SolveEqMotionD[ Lag ];
  Lag = SolveEqMotionF[ Lag ];
\end{verbatim}
In addition, all Weyl fermions must be replaced in terms of their four-component
counterparts,
\begin{verbatim}
  Colourb = Colour;
  Lag = Lag /. { Tb[a_,i_,j_]->-T[a,j,i] };
  Lag = ExpandIndices[ Lag , FlavorExpand -> {SU2W, SU2D} ];
  Lag = WeylToDirac[ Lag ];
\end{verbatim}
The set of commands above first maps the matrices of the anti-fundamental
representation of $SU(3)_c$ to those related to the fundamental one
(see section \ref{sec:theolag}), and then performs
an expansion of all the indices, fundamental or adjoint, related to the
$SU(2)_L$ group. This procedure forces the $SU(2)_L$ field rotations from the
gauge basis to the mass basis, which is necessary for the function {\tt
WeylToDirac} to correctly perform the translation to Dirac and Majorana fermions
\cite{Duhr:2011se}. Let us note that in the current version of \feynrules, the
indices \texttt{SU2W}, \texttt{SU2D} and \texttt{Colourb} are non standard.
Therefore, the manipulations above have to be performed for any model. However,
in future versions of the code, the issue of automating this step will be
addressed.

\subsection{The Minimal $R$-symmetric Supersymmetric Standard Model}
\label{sec:mrssm}
\subsubsection{Theoretical framework} \label{sec:mrssmth}
The supersymmetry algebra naturally contains a continuous $R$-symmetry and there
is a large class of mechanisms leading to supersymmetry breaking without
breaking this $R$-symmetry \cite{Intriligator:2006dd,Intriligator:2007py}.
Among the major consequences of the conservation of this symmetry, 
Majorana gaugino masses are forbidden. 
Therefore, phenomenologically viable models requires the pairing of each gaugino
with the fermionic component of a new chiral superfield to form a massive 
Dirac fermion \cite{Polchinski:1982an,Dine:1992yw,Fox:2002bu}.

The preservation of the $R$-symmetry also forbids bilinear
Higgs mixing term of the superpotential and  soft-supersymmetry breaking
trilinear scalar interactions. Therefore, phenomenological viability requires
the introduction of another set of chiral superfields, the 
$R$-partners of the Higgs superfields. Mixing terms between the Higgs
and these new superfields can be included in the superpotential. Consequently,
this renders the Higgsino fields massive and restores agreement with the
experimental non-observation of a massless Higgsino field.

Finally, an unbroken $R$-symmetry also ensures that most of the
dangerous flavor-changing operators which could appear in generic
supersymmetric theories are naturally forbidden.

Starting from the gauge group $G_{MSSM}
\equiv SU(3)_c \times SU(2)_L \times U(1)_Y$ and the 
MSSM superfield content presented in Tables \ref{tab:gauge} and
\ref{tab:chiral}, we add the three chiral superfields $\Phi_B$, $\Phi_W$ and
$\Phi_G$
to the theory. They lie in the adjoint representation of
the relevant gauge group,
\be
  \Phi_B = ({\utilde{\bf 1}}, {\utilde{\bf 1}}, 0) \ , \qquad  
  \Phi_W = ({\utilde{\bf 1}}, {\utilde{\bf 3}}, 0) \ , \qquad
  \Phi_G = ({\utilde{\bf 8}}, {\utilde{\bf 1}}, 0) \ .
\ee
Let us note that they form, together with the vector superfields $V_B$, $V_W$
and $V_G$,
three complete vector representations of the $N=2$ supersymmetric algebra.

The Higgsino fields can be rendered massive by mixing the two Higgs chiral superfields $H_D$
and $H_U$ with their $R$-partners, the chiral superfields $R_D$ and $R_U$, 
\be
  R_D = ({\utilde{\bf 1}}, {\utilde{\bf 2}}, \frac12) \ , \qquad  
  R_U = ({\utilde{\bf 1}}, {\utilde{\bf 2}}, -\frac12) \  .
\ee
The notations related to the component fields of these five new superfields are
shown in Table \ref{tab:mrssm}.

\renewcommand{\arraystretch}{1.3}
\begin{table}[h]
  \tbl{$R$-partners of the vector and Higgs superfields in
    the minimal $R$-symmetric supersymmetric model. They are given together with
    their scalar and fermionic components, as well as with their
    representations under the $SU(3)_c \times SU(2)_L \times
    U(1)_Y$ gauge group.}
  {\begin{tabular}{|c||c|c|c|c|}
    \hline
    Superfield & Fermion & Scalar & Representation \\ \hline\hline 
    $\Phi_B$ & $\widetilde B'$ & $\sigma_B$ & 
       $({\utilde{\bf 1}}, {\utilde{\bf 1}}, 0)$\\
    $\Phi_Y$ & $\widetilde W'$ & $\sigma_W$ & 
       $({\utilde{\bf 1}}, {\utilde{\bf 3}}, 0)$\\
    $\Phi_G$ & $\widetilde g'$ & $\sigma_G$ & 
       $({\utilde{\bf 8}}, {\utilde{\bf 1}}, 0)$\\
    \hline 
    \multirow{4}{*}{$R_U$}&&&\\
       &$\widetilde R_u = \bpm \widetilde R_u^0 \\ \widetilde R_u^-\epm$&
       $R_u = \bpm R_u^0 \\ R_u^- \epm$ &         
        $({\utilde{\bf 1}}, {\utilde{\bf 2}}, -\frac12)$ \\&&&\\
    \multirow{4}{*}{$R_D$}&&&\\
       &$\widetilde R_d = \bpm \widetilde R_d^+ \\ \widetilde R_d^0\epm$&
       $R_d = \bpm R_d^+ \\ R_d^0 \epm$ &         
        $({\utilde{\bf 1}}, {\utilde{\bf 2}}, \frac12)$ \\&&&\\
   \hline
   \end{tabular}\label{tab:mrssm}}
\end{table}
\renewcommand{\arraystretch}{1}

Kinetic and gauge interaction terms for the new superfields $\Phi_B$, $\Phi_W$
and $\Phi_G$ are given by the canonical K\"ahler potential of Eq.\
\eqref{eq:renormsusy},
\be
  \lag_K = 
    \int \d^2\theta\d^2\thetabar\ \bigg[ \Phi_B^\dag \Phi_B +  
      \Phi_W^\dag e^{-2 g_w \tilde V_W} \Phi_W +  
      \Phi_G^\dag e^{-2 g_s \tilde V_G} \Phi_G \bigg] \ ,
\label{eq:sglkin}\ee
where $\tilde V_W = V_{W^k} T_k$ and $\tilde V_G = V_{G^a} T_a$. The vector
superfields $V_{W^k}$ and $V_{G^a}$ have been defined in Eq.\
\eqref{eq:VSFmssm}, and, in this case, the matrices $T_k$ and $T_a$ are taken as 
the representation matrices of the $SU(2)$ and $SU(3)$ algebra in the adjoint
representation. The Lagrangian of
Eq.\ \eqref{eq:sglkin} contains couplings of the new scalar adjoint 
$\sigma$-fields to a single gauge boson and to pairs of gauge bosons through the usual
covariant derivatives. The gauge-invariant kinetic terms of the $R$-superfields
are given as in the second term of Eq.\ \eqref{eq:lrpv}, the representation
matrices of $SU(2)$ being taken in the fundamental representation.

Interactions among the chiral superfields of the theory is given by the
superpotential
\be\bsp
  W(\Phi) =&\  
    ({\bf y^u})_{ij} U_R^i Q_L^j \!\cdot\! H_U - 
    ({\bf y^d})_{ij} D_R^i Q_L^j \!\cdot\! H_D - 
    ({\bf y^e})_{ij} E_R^i L_L^j \!\cdot\! H_D \\ &\ 
   + \sum_{i=U,D} \bigg[ \lambda^B_i H_i \Phi_B Y R_i 
    + \lambda^W_i H_i \Phi_{W^k} \frac{\sigma_k}{2} R_i  + \mu_i H_i\!\cdot\!R_i \bigg] \ .
\esp\label{eq:Wmrssm}\ee
In the expression above, ${\bf y^u}$, ${\bf y^d}$ and ${\bf y^l}$ denote the
$3\times3$ Yukawa matrices in generation space and these terms are also present
in the MSSM. In contrast, the other terms are related to the $R$-partner
superfields. The parameters $\mu_i$ are the ($R$-)Higgs off-diagonal mass-mixing
parameters and the dot
products again stand for $SU(2)$ invariant products. The $\lambda$-couplings are the
trilinear couplings of the $R$-partners of the $U(1)_Y$ and $SU(2)_L$ vector
superfields with the ($R$-)Higgs superfields.

Soft-supersymmetry breaking originates from hidden sector spurions which
preserve the $R$-symmetry. In the most general version of the model, both
$F$-type and $D$-type supersymmetry breaking terms are 
allowed. In the first case, a spurion chiral superfield $X$ gets a
vacuum expectation value $\langle X \rangle = \theta\!\cdot\!\theta v_F$ whilst
in the second case, a spinorial field strength tensor $W'_\alpha$ gets a vacuum
expectation value $\langle W'_\alpha\rangle = \sqrt{2} \theta_\alpha v_D$, the
quantities $v_F$ and $v_D$ being related to the order of magnitude $M_{\rm
SUSY}$ of the supersymmetric
masses. 

Dirac gaugino masses arise through dimension-five
operator generating mixing between the fermionic component of the chiral
superfield $\Phi_k$ and the gaugino component of the vector superfield $V_k$,
with $k=B,W,G$,
\be
   \lag_{\rm soft,1} =
     \sum_{k=B,W,G} \frac{1}{2 g_k} m_1^k \int \d^2\theta\ \frac{W^{\prime\alpha}}{M_{\rm SUSY}}
        W_{k\alpha} \Phi_k \ .
\label{eq:lbrk1}\ee
This involves a $D$-type spurion, $W_k$ being the superfields strength tensor
associated to the vector superfield $V_k$ and $m_1^k$ the corresponding mass
parameter. The overall normalization factor $1/(2 g_k)$ depending on the 
gauge coupling constant ensures a correct normalization of the Dirac mass term.

The $F$-type spurion allows to write down mass terms for the
scalar sector of the theory,
\be
  \lag_{\rm soft,2} =\sum_{\Phi = \text{chiral superfields}} \
     \int \d^2 \theta\d^2\thetabar\ m_\Phi^2 \frac{X X^\dag}{M_{\rm SUSY}^2} \Phi^\dag
     \Phi \ .
\label{eq:lbrk2}\ee
The sum runs over the chiral superfields included in both Table \ref{tab:chiral}
and Table \ref{tab:mrssm}. However, both $F$-type and $D$-type spurions allow to
generate additional mass terms for the scalar
adjoint $\sigma$-fields through the dimension-six operators 
\be\bsp
   \lag_{\rm soft,3}  = \sum_{k=B,W,G} \int \d^2 \theta\d^2\thetabar\ \bigg[ 
      (m_2^k)^2 \frac{X X^\dag}{M_{\rm SUSY}^2} + (m_3^k)^2 \frac{W'\!\cdot\!
      W'}{M_{\rm SUSY}^2} \bigg] {\rm Tr} \big[\Phi_k \Phi_k
     \big] \ .
\esp\label{eq:lbrk3}\ee
Assuming real soft-supersymmetry breaking masses, the two real degrees of
freedom included in each of the complex $\sigma$-scalar fields do not mix.
Consequently,
we have purely scalar or purely pseudoscalar mass-eigenstates. For the sake of
simplicity, this hypothesis is adopted in this work and the pseudoscalar
state are neglected.

Finally, the soft-supersymmetry breaking Lagrangian also contains bilinear
$(R-)$Higgs interactions driven by the $F$-type spurion,
\be
  \lag_{\rm soft,4} = \int \d^2 \theta\d^2\thetabar\  \frac{X X^\dag}{M_{\rm
    SUSY}^2} \bigg[ B_U H_U\!\cdot\!R_U + B_D H_D\!\cdot\!R_D + B
    H_u \!\cdot\! H_D \bigg] \ ,
\label{eq:lbrk4}\ee
where $B$, $B_U$ and $B_D$ are the corresponding mixing strengths.

More importantly for hadron-collider phenomenology, the new scalar fields also
couple singly to a pair of quarks and to a pair of gluons through loop-diagrams
involving squarks, gluinos, neutralinos and charginos. Enhanced by the relative
magnitude (at the weak scale) of the strong coupling, we only keep the dominant
interactions involving a sgluon field $\sigma_G$. They can be described by
the effective Lagrangian
\be\bsp
  \lag_{\rm eff} = &\ \sigma_G^a \Big [ 
     (\lambda_L)^f{}_{f'} \bar \Psi_{qf} P_L T_a \Psi_q^{f'}  +  
     (\lambda_R)^f{}_{f'} \bar \Psi_{qf} P_R T_a \Psi_q^{f'}\Big]  \\ &\ + 
    \lambda_g d^{abc} \sigma_{Ga} G_{\mu\nu b} G^{\mu\nu}{}_c
    + {\rm h.c.} \ .
\esp\label{eq:Leff}\ee 
In this last equation, we have introduced  $\Psi_q^f$ as the Dirac field
associated to a quark of flavor $f$, 
$G_{\mu\nu b}$ as the gluon field strength tensor and $d^{abc}$ as the $SU(3)$
symmetric tensor. The strengths of the interactions included in the effective
Lagrangian $\lag_{\rm eff}$, \ie, the parameters $\lambda_L$, $\lambda_R$ and
$\lambda_g$, are taken as free parameters even if in principle, they
fully depend on the particle spectrum and sfermion mixings.

\subsubsection{Implementation in \feynrules}
As in Section \ref{sec:rpv}, the implementation in \feynrules\ 
of the model described above can
be performed in a very efficient way starting from the
existing implementation of the MSSM. However, unlike the
$R$-parity violating model, all the novelties cannot be included in an
additional file only. The declaration of the gluino and neutralino fields must indeed be
modified, since they are now Dirac instead of Majorana fermions, and this
requires a modification of the MSSM implementation. Therefore, we start from a
copy of the file {\tt mssm.fr} and include all the modifications 
presented in Section \ref{sec:mrssmth}.

The implementation of the five new chiral superfields as well as the one of all 
the model free parameters strictly follow the
rules presented in Ref.\ \refcite{Christensen:2008py} and Ref.\
\refcite{Duhr:2011se}. Whilst the implementation of the parameters and the one
of the $R$-Higgs superfields is immediate, let us take the example of the
$\Phi_G$ 
superfield to briefly comment on the implementation of the $R$-partners of the 
vector superfields of the MSSM. A possible set of \mathematica\ commands for
the declaration of the superfield $\Phi_G$ reads
\begin{verbatim}
  CSF[9] == {
    ClassName -> SGL,
    Chirality -> Left,
    Scalar    -> sigG,
    Weyl      -> gopw,
    Indices   -> { Index[Gluon] } 
  }
\end{verbatim}
where we have assigned the symbols {\tt sigG} and {\tt gopw} to the sgluon
$\sigma_G$ and the gluino $\tilde g'$ field, respectively. For the
implementation of the component fields, we have,
\begin{verbatim}
   S[100] == { 
     ClassName     -> sigG,
     Unphysical    -> True, 
     SelfConjugate -> False,
     Indices       -> { Index[Gluon] },
     Definitions   -> { sigG[a_] -> sig1[aa] /Sqrt[2] ] } 
   },
   S[101] == { 
     ClassName     -> sig1, 
     SelfConjugate -> True,
     Indices       -> { Index[Gluon] },
     Mass          -> Msig1,
     Width         -> Wsig1
   },
   W[100]== { 
     ClassName     -> gopw,
     Unphysical    -> True, 
     Chirality     -> Left,
     SelfConjugate -> False,
     Indices       -> {Index[Gluon]}
   }
\end{verbatim}
Before moving on, let us comment on the way used to implement
the mixing of the real degrees of freedom included in the complex sgluon field. 
Whilst the symbol {\tt sigG} represents the complex scalar field which is the
unphysical component field of the chiral superfield $\Phi_G$, 
the symbol  {\tt sig1} is the real
scalar (and not pseudoscalar) degree of freedom included in $\sigma_G$. 
The pseudoscalar state has indeed been decoupled and removed from the mixing
relation implemented in the option {\tt Definitions} of the class {\tt sigG}. The 
$\widetilde g'$ field has been associated to the symbol {\tt gopw}. Denoting by
{\tt gow} the gluino $\widetilde g$ two-component fermion embedded in $V_G$, the
two fields $\widetilde g$ and $\widetilde g'$ can be related 
to the (Dirac) four-component gluino field through the option
\begin{verbatim}
  WeylComponents -> {gow, gpowbar}
\end{verbatim}
of the gluino instance of the particle class. 

We now turn to the $R$-symmetric Lagrangian presented in Section
\ref{sec:mrssmth}. All the gauge-invariant kinetic terms which are given in Eq.\
\eqref{eq:lrpv} and Eq.\ \eqref{eq:sglkin} are automatically computed by issuing
the commands
 \begin{verbatim}
  LagKin = Theta2Thetabar2Component[ CSFKineticTerms[ ] ] + 
    Theta2Component[ VSFKineticTerms[ ] ] +
    Thetabar2Component[ VSFKineticTerms[ ] ];
\end{verbatim}
The superpotential can also be immediately implemented from Eq.\
\eqref{eq:Wmrssm}, translating the textbook expression into a \mathematica\
declaration of a variable {\tt SuperW},
\begin{verbatim}
   SuperW =  ...
     -luB/2 (HU[1] PhiB RU[2] - HU[2] PhiB RU[1]) + 
      ldB/2 (HD[1] PhiB RD[2] - HD[2] PhiB RD[1]) + 
     luW PhiW[a] (HU[1] Ta[a,2,i] RU[i] - HU[2] Ta[a,1,i] RU[i]) + 
     ldW PhiW[a] (HD[1] Ta[a,2,i] RD[i] - HD[2] Ta[a,1,i] RD[i]) + 
     MUu (HU[1] RU[2] - HU[2] RU[1]) + 
     MUd (HD[1] RD[2] - HD[2] RD[1]) ];
\end{verbatim} 
where the dots stand for the trilinear Yukawa interactions identical as in the
MSSM
and then omitted for brevity \cite{Duhr:2011se}. In the implementation above,
the $SU(2)_L$ contractions have been expanded, {\tt Ta} are the symbols
representing the fundamental representation matrices of $SU(2)_L$, {\tt HU},
{\tt HD}, {\tt RU} and {\tt RD} are the names of the classes associated to the
($R$-)Higgs superfields,
and {\tt PhiB} and {\tt PhiW} to the superfields $\Phi_B$ and $\Phi_W$. The
superpotential $\lambda$-parameters of the superpotential are denoted by {\tt
luB}, {\tt ldB}, {\tt luW} and {\tt ldW} whilst the $\mu$-parameters are
given as {\tt MUu} and {\tt MUd}. The associated Lagrangian can then be computed
as
\begin{verbatim}
  Lag = Theta2Component[SuperW] + Thetabar2Component[HC[SuperW]];
\end{verbatim}
Finally, the soft-supersymmetry breaking Lagrangian split between  
Eq.\ \eqref{eq:lbrk1}, Eq.\ \eqref{eq:lbrk2}, Eq.\ \eqref{eq:lbrk3} and Eq.\
\eqref{eq:lbrk4} can also be implemented directly within the superspace module
of \feynrules. As an example, we take the gluino Dirac mass term implemented in
the variable {\tt mglno} which could be
included in the model file as
\begin{verbatim}  
    mglno  = MG1/(2 gs) Ueps[be,al] * 
      nc[theta[al], SuperfieldStrengthL[GSF, be, a], PhiG[a]]];
    LSoft = Theta2Component[mglno] + Thetabar2Component[HC[mglno];
\end{verbatim}
where {\tt MG1} is the symbol associated to the product of the soft
supersymmetry-breaking mass by the vacuum expectation value of the
spurion superfield strength tensor $W'$, \ie, $m_1^G v_D$. For the conventions
on the rank-two fully antisymmetric tensor {\tt UEps} and on the {\tt nc}
environment, we refer to Ref.\ \refcite{Duhr:2011se}.

\section{From \feynrules\ to \madgraph} \label{sec:UFOMG}
\subsection{General features}

Among the whole set of existing automated Monte Carlo tools allowing to address
phenomenological studies of high-energy physics processes, most all of them
contain restrictions on the color and/or Lorentz structures which can
appear in the interaction vertices of any model. While all the structures
included in the Standard Model and the MSSM are generally allowed,
vertices with non-standard color and/or Lorentz structures are most of the time
not supported and must be discarded from the model implementations.
As a consequence, this constrains beyond the Standard Model phenomenological
explorations which could be performed with the help of an automated Monte Carlo
tool. One is thus forced to
compute squared matrix element in a non-automated fashion, \eg, by hand, and
implement the results into non-automated tools such as \herwig\
\cite{Corcella:2000bw,Bahr:2008pv} or \pythia\
\cite{Sjostrand:2006za,Sjostrand:2007gs}. Going beyond two-to-two scattering
processes is thus a rather tedious task.

In this work, we present a way to overcome those limitations with the use of
the \madgraph\ 5 matrix element generator \cite{Alwall:2011uj} and the UFO model 
format
\cite{Degrande:2011ua}. We focus on the non-trivial color structures appearing 
in the Lagrangians of the theories presented in Section \ref{sec:feynrules},
namely the rank-three fully antisymmetric tensor in color space $\epsilon_{\ell
m n}$, where $\ell$, $m$ and $n$ are (anti)fundamental color indices, and the
symmetric tensor of $SU(3)$, $d^{abc}$, where $a$, $b$ and $c$ are adjoint color
indices.

The Monte Carlo event generator \madgraph\ 5 allows for the automated
generation of tree-level matrix elements associated to any scattering 
process in a very efficient way. In particular, \madgraph\ 5 is thus a
suitable tool to simulate final state signatures
such as those produced in the high energy 
proton-proton collisions occurring at the LHC.
The task of the user consists mainly in specifying the process of interest in
terms of initial and final state particles, the collision basic information 
(such as the energy of the colliding beams), a set of basic cuts related to 
the analysis aimed to be performed and, of course, the particle physics model 
under consideration, renormalizable or not.

The model library of \madgraph\ 5 is built upon the \feynrules\ model database.
Information is passed from \feynrules\ to \madgraph\ through the universal
output format of \feynrules, dubbed the UFO. This format has been designed to
overcome the above-mentioned restrictions on the allowed Lorentz and color
structures for the interaction vertices at the level of the Monte Carlo tools.
The key features are flexibility and
modularity through the use of \python\ classes and objects to represent
particles, parameters and vertices, saving thus the model information in an
abstract fashion, also independent from any Monte Carlo program. Consequently,
it is universal in the sense that it is not tied to any specific matrix element
generator. Presently, two generators, \madgraph\ and \gosam, are currently using
it as their standard 
model format, and the future version of \herwig++ will also be based on the UFO
format. 

In order to translate a \feynrules\ model implementation, such as one of those
described in Section \ref{sec:feynrules}, an automated interface has been
included within the public version of the \feynrules\ program. It can be called
 by typing, in a \mathematica\ session, the command
\begin{verbatim}
   WriteUFO[ Lag ] ; 
\end{verbatim}
where the variable {\tt Lag} contains the Lagrangian of the considered model,
expressed in terms of the usual scalar fields, (four-component) fermions, gauge
bosons (together with the corresponding ghost fields, if relevant) 
and tensorial fields of particle physics. Let
us note that all the states are assumed to be rotated to the mass eigenbasis. 
Once issued, the {\tt WriteUFO}
function internally calls the \feynrules\ core function {\tt FeynmanRules} in
order to compute all the vertices associated to the Lagrangian {\tt Lag}. They
are then expanded into a color $\otimes$ spin basis,
\be\label{eq:ufovert}
 \begin{cal}V\end{cal}^{a_1\ldots a_n, \ell_1\ldots\ell_n}(p_1,\ldots,p_n) =
    \sum_{i,j}C_i^{a_1\ldots a_n}\,G_{ij}\,L_j^{\ell_1\ldots\ell_n}(p_1,\ldots,p_n)
     \ , 
\ee
where the variables $p_i$ denote the particle momenta and $G_{ij}$ the coupling
strengths. The quantities $C_i^{a_1\ldots a_n}$ and
$L_j^{\ell_1\ldots\ell_n}(p_1,\ldots,p_n)$ are tensors in color and spin space,
respectively. These tensors act as a basis in the color $\otimes$ spin space and
could be used by several vertices, reducing by this way possible redundancies in
a model implementation. The coupling strengths of a specific vertex are
therefore the coordinates of the vertex in this basis. Following this
structure, vertices are implemented in the UFO format by using
several different \python\ objects and their attributes to 
represent the vertices themselves, their
Lorentz and color structures and the coupling strengths.

Once all the Lagrangian vertices are decomposed as in Eq.\ \eqref{eq:ufovert},
the UFO interface writes a set of \python\ files stored in a single directory. These
files contain all the model information, from the definition of the particle
content and the parameters as instances of the generic UFO
particle and parameter class,
respectively, to the implementation of the (factorized) vertices as instances of
the vertex class. 

In order to use the produced UFO model with \madgraph, the outputted files must
be copied into 
the \texttt{models} directory of \madgraph\ so that they can be used for event generation
as any other built-in model. 

As soon as a process is specified by the user, \madgraph\ 5 internally calls the 
\aloha\ package \cite{deAquino:2011ub} which generates the subroutines, based on
the \helas\ library \cite{Murayama:1992gi,Hagiwara:2008jb,Hagiwara:2010pi,%
Mawatari:2011jy}, necessary to compute the helicity amplitudes
related to the process under consideration from the UFO. This allows for an
efficient evaluation of the associated squared matrix element. Indeed, helicity
amplitudes include helicity wavefunctions corresponding to specific
substructures included in a given Feynman diagram, which can be further reused
within different diagrams. 

Supersymmetric theories contain, in their most general form, more than
several thousands of vertices. A large class of vertices are however
vanishing in scenarios of phenomenological interests
at colliders. As an example, let us take the sector of the up-type
squarks (the six scalar partners of the three generations of left-handed and
right-handed up-type quarks). Since after
electroweak symmetry breaking, all particles with the same spin, color
representation and electric charge mix, the six up-type squark gauge-eigenstates
undergo a $6\times 6$ mixing. Consequently, the scalar potential contains
${\cal O}(1000)$ four scalar interactions among the up-type squark
mass-eigenstates which violate flavor most of the time. However, the
flavor-violating entries of the associated mixing matrix are, in most 
phenomenologically viable benchmark scenarios, vanishing, which renders 
a large part of the ${\cal O}(1000)$ interaction vertices equal to
zero and irrelevant. 

The presence of zero vertices in UFO model files, or more generally into model
files for any matrix element generator, considerably slows down event
generation, since they must be loaded in the computer memory on run time and 
diagrams with a vanishing contribution 
are effectively generated. Therefore, it may be suitable to remove
these vertices from the UFO model files to speed up the evaluation of the matrix
elements, making it much more efficient. 
This task can be done at the \feynrules-level with the help of the
{\tt WriteRestrictionFile} and {\tt LoadRestriction} commands.

Firstly, the 
numerical values of all the model parameters, \ie, a full supersymmetric
spectrum, must be loaded in \feynrules. Since the instances of the
parameter class included in \feynrules\ models are organized following a
structure inspired by the Supersymmetry Les Houches Accord (SLHA),
the numerical initialization of the model parameters can be achieved
by loading a SLHA file directly into the \mathematica\ session, 
\begin{verbatim}
  ReadLHAFile[Input -> "susy.dat"];
\end{verbatim}
where {\tt susy.dat} is the filename of the spectrum provided by, \eg, one of
the dedicated existing supersymmetric spectrum generators. 
The detection of the vanishing parameters is performed by issuing the following
sequence of \mathematica\ commands,
\begin{verbatim}
  WriteRestrictionFile[ ]; 
  LoadRestriction["ZeroValues.rst"];
\end{verbatim}
The {\tt WriteRestrictionFile} function asks \feynrules\ to scan over the whole
set of internal and external
parameters and write, in a file named {\tt ZeroValues.rst}, a list of
\mathematica\ replacement rules mapping all the vanishing parameters to zero.
The {\tt LoadRestriction} function allows to read this file and load the list of
rules
into \feynrules\ so that it could be used, at a later stage, by either the UFO
interface or by any other \feynrules\ interface to a Monte Carlo program.
When called, the interfaces apply this set of replacement rules to each vertex
derived by the function {\tt FeynmanRules} before translating it to be 
written to the output
files. The zero contributions are hence immediately simplified and dropped. 
If after the mapping, a full vertex is numerically evaluated to zero, it is
ignored and not written at all in the generated output files.

After this optimization, it can be noted that among all the remaining
interactions, hundreds of vertices consist in four scalar interactions which
are, for tree-level computations, in general phenomenologically less relevant.
Therefore, again for the sake of efficiency at the Monte Carlo tool level,
it is useful not to include those vertices. This task can
be done through the {\tt Exclude4Scalars} option of the \feynrules\ interfaces.
In the UFO case, one would have to issue the \mathematica\ command
\begin{verbatim}
 WriteUFO[ Lag, Exclude4Scalars -> True ];
\end{verbatim}

Let us note that one must keep in mind that the UFO model files generated in the
optimized way illustrated above are not fully general and highly dependent on
the considered benchmark scenario (defined here in the file {\tt susy.dat}),
even if at the \feynrules\ level, the model implementation is as general as
possible.

\section{Benchmark scenarios and processes of interest}\label{sec:bench}
\subsection{The $R$-parity violating Minimal Supersymmetric Standard Model}
The ATLAS and CMS experiments are currently setting impressive limits on the
masses of the supersymmetric particles and excluding a significant part of the
supersymmetric parameter space \cite{ATLAS,CMS}. In order to reinterpret the
experimental results in terms of as many different manifestations of
supersymmetry as possible, specific benchmark scenarios have been recently 
proposed \cite{AbdusSalam:2011fc}. Even if a large emphasis is put on the 
constrained version of the MSSM, the $R$-parity
violating case is also addressed. We therefore adopt as a benchmark
scenario for our $R$-parity violating supersymmetric exploration one of the
proposed points, along a line in the $R$-parity violating MSSM parameter
space dubbed the `RPV3-line'\cite{AbdusSalam:2011fc}.

This line is inspired by the constrained MSSM. Each benchmark lying on it is
thus defined by four free parameters defined at the grand unification scale and
the sign of the bilinear Higgs mixing superpotential $\mu$-parameter. To this
restricted set of supersymmetric input parameters, one must supplement the value
of one of the $\lpp$ coupling, assuming thus the single coupling dominance
hypothesis \cite{Dimopoulos:1988jw,Barger:1989rk} and neglecting any other
$\lpp$, together with all the $\lambda$, $\lambda'$, $T$, $T'$ and
$T''$ parameters. This line is thus perfectly suitable to probe the exotic color
structures included in the superpotential part of the Lagrangian.

We fix the universal scalar mass
$m_0=100$ GeV, the universal gaugino mass $m_{1/2}=400$ GeV and the universal
trilinear coupling $A_0=0$ GeV. The ratio of the vacuum expectation
values of the neutral component of the two Higgs doublets is taken as
$\tan\beta=10$, whilst the $\mu$-parameter is chosen positive. 

For the Standard Model sector, we fix the top quark
pole mass to $m_t = 173.2$ GeV \cite{Lancaster:2011wr}, 
the bottom quark mass to $m_b(m_b) = 4.2$ GeV and the $Z$-boson mass to $m_Z =
91.1876$ GeV. The Fermi constant has been taken as $G_F = 1.16637 \times
10^{-5}$ GeV$^{-2}$, and the strong and electromagnetic coupling constants at the
$Z$-pole as $\alpha_s(m_Z) = 0.1176$ and $\alpha(m_Z)^{-1} = 127.934$, according
to the Particle Data Group Review \cite{Nakamura:2010zzi}.

There exists a wide range of experimental measurements constraining the magnitude
of the baryon-number violating $\lpp$ parameters, such as data related to, \eg, 
$K-\bar K$
systems \cite{Barbieri:1985ty,Abel:1996qj,Slavich:2000xm}, neutron
electric dipole moments \cite{Barbieri:1985ty}, rare hadronic $B$-decays  
\cite{Chakraverty:2000df,BarShalom:2002sv}, nucleon-antinucleon
oscillations, as well as to double
\cite{Zwirner:1984is,Dimopoulos:1987rk,Hinchliffe:1992ad} and 
single nucleon decays \cite{Chang:1996sw,Choi:1998ak,Bhattacharyya:1998dt}.
However, the most restrictive bounds on the $\lpp$ parameters are deduced from 
cosmological and astrophysical data with the observed flux of antiproton cosmic
rays \cite{Baltz:1997ar}. All these limits are nevertheless not applicable on
the $\lpp_{3jk}$ couplings, related to the top (s)quarks, for supersymmetric
scenarios where the lightest supersymmetric particle is lighter than the top
quark. In this case, the $\lpp_{3jk}$ parameters are left almost unconstrained,
which is particularly appealing for LHC physics. We choose $\lambda_{312} =
0.2$, since with our setup, at it is presented below, the lightest
supersymmetric particle is lighter than the top quark.

The input parameters for the chosen benchmark point are summarized in Table
\ref{tab:rpvbenchmark}, where we remind that the supersymmetric parameters are
defined at the grand unification scale. 

\renewcommand{\arraystretch}{1.3}
\begin{table}[h]
  \tbl{Input parameters associated to the chosen benchmark scenario in
    the context of the $R$-parity violating MSSM. All other $\lpp$-parameters,
    together with the $\lambda$, $\lambda'$, $T$, $T'$ and $T''$ parameters
    are taken equal to zero.}
  {\begin{tabular}{|cccccc|}
    \hline $m_t$ [GeV] & $m_b$ [GeV] &$m_Z$ [GeV]& $G_F$ [GeV$^{-2}$] &
      $\alpha_s(m_Z)$& $\alpha(m_Z)^{-1}$  \\
    \hline
    173.2 & 4.2 & 91.1876 & $1.16637 \times 10^{-5}$ & 0.1176 & 127.934 \\ 
   \hline\hline
    $m_0$ [GeV]  & $m_{1/2}$ [GeV] & $A_0$ [GeV] & 
      $\tan\beta$ & $\text{sign}(\mu)$ &
      $\lambda''_{312}$ \\
    \hline
    100 & 400 & 0 & 10 & $>0$ & 0.2 \\ 
    \hline 
   \end{tabular} \label{tab:rpvbenchmark} }
\end{table}
\renewcommand{\arraystretch}{1.}

The soft supersymmetry-breaking masses and interaction strengths at the
electroweak scale are obtained through renormalization group running using the
\spheno\ 3 package \cite{Porod:2003um,Porod:2011nf}, which solves the
renormalization group equations numerically to two-loop order. This program then
extracts the particle spectrum and mixings at the two-loop level for the Higgs
sector and at the one-loop level for all the other particles. 

In the sector of the electroweak superpartners, the sleptons are fairly light,
with masses of ${\cal O}(200-300)$ GeV, and the neutralino and chargino masses
range from 160 GeV for the lightest neutralino, being the lightest
supersymmetric particle, to 550 GeV for the heavier
states. Let us emphasize that since the lightest neutralino is lighter than the
top quark, the conditions to evade the experimental bounds on the $\lpp$
parameters are fulfilled. The choice of $\lpp_{312}=0.2$ is thus justified.
In contrast, in the strong sector, the squark masses are a larger,
even if still rather modest, ranging from 650 GeV to 875 GeV, whilst the gluino
is heavier than all the squarks with a mass of 900 GeV.

\begin{figure}[t]
\vspace{6cm}\hspace{-2cm}
\centerline{\psfig{file=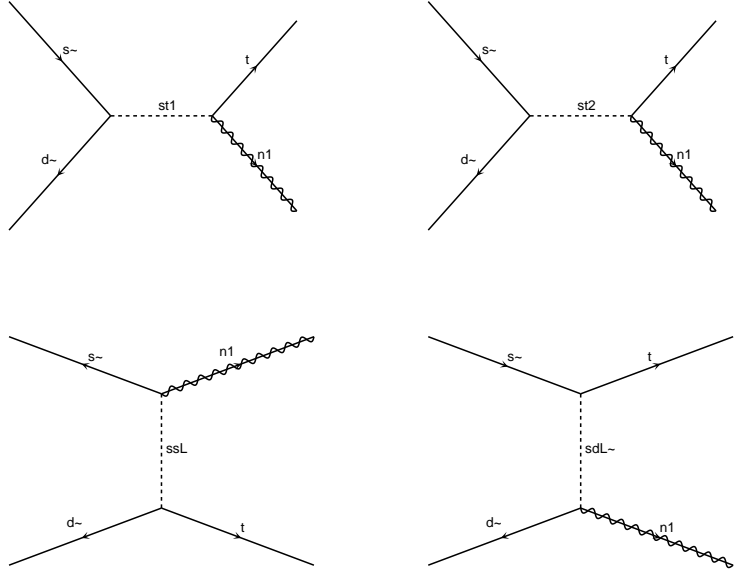,width=.75\columnwidth}}
\vspace*{8pt}
\vspace{-6.5cm}
\caption{Feynman diagrams associated to $R$-parity violating monotop
production as generated by \madgraph\ 5. In the notations above, the 
label \texttt{n1} stands for the lightest neutralino, \texttt{s$\sim$} and
\texttt{d$\sim$} for strange and down antiquarks and \texttt{t} for a top
quark. Attached to the internal lines, \texttt{st1} and \texttt{st2} represent
the two top squarks, \texttt{ssL} the left-handed strange squark and
\texttt{sdL$\sim$} the left-handed down antisquark.
\protect\label{fig:monotopdiag}}
\end{figure}

One can observe that in the scenario depicted above, top squarks and antisquarks 
can be singly produced at the LHC with a sensible rate from an initial
associated pair of down and strange (anti)quarks. Employing the CTEQ6L1
set of parton densities \cite{Pumplin:2002vw} and for a center-of-mass energy of
7 TeV, the corresponding total hadronic cross section indeed reach the level of
about 2 pb. Whereas top squarks dominantly decay to dijets, their observation in
this channel is made difficult by the huge QCD background. In contrast,
the produced (anti)squark can decay to an associated pair of a top quark and a
lightest neutralino with a branching ratio of 15\%.

As $R$-parity is violated through baryon-number violating operators, 
one can expect a multijet signature after the decay of both the top
quark and the neutralino. However, in the considered scenario, the only 
possible three-body decay of the lightest neutralino, via a $\lpp_{3jk}$
interaction, is not kinematically allowed and the only remaining option goes
through a four-body decay. As a consequence, the lightest neutralino has a long
lifetime and decays far outside the detector \cite{Allanach:1999bf}. Therefore,
the missing energy signature, typical from $R$-parity conserving supersymmetric
scenarios, is still a convenient observable. This will be investigated in
Section \ref{sec:pheno} where we carefully study the $R$-parity violating
production of a single top quark in associated with missing energy, this
signature being dubbed as a monotop signature \cite{Andrea:2011ws}. The
corresponding Feynman diagrams, as generated by \madgraph\ 5, are presented in
Figure \ref{fig:monotopdiag}.

\subsection{The Minimal $R$-symmetric supersymmetric model}
\label{sec:benchmrssm}
As stated above, most of the results of the ATLAS and CMS experiments are
derived under the hypothesis of the constrained MSSM. Several reinterpretations
exist in the context of popular models such as, \eg, the MSSM where
supersymmetry is broken through gauge interactions, the so-called
phenomenological MSSM or the Next-to-Minimal Supersymmetric Standard Model.
However, there are alternative realizations of supersymmetry with highly
different properties. Therefore, their investigation deserves
dedicated studies in order to prepare the reinterpretation of the data.
The minimal $R$-symmetric supersymmetric model is one of
such examples, since it predicts, among others, the existence of a new
color-octet scalar field (dubbed the sgluon) which is not present in more
conventional experimentally covered supersymmetric theories. In this Section, we
address the phenomenology of such as field, from its production at the LHC to
its decay and signature within a detector.

In the model presented in Section \ref{sec:mrssm}, sgluons can be
either singly or pair-produced in hadron collisions at high energies 
from initial states of quarks and gluons. Since the sgluon field belongs to the
Standard Model sector of supersymmetric theories\footnote{The sgluon
field carries a positive $R$-parity quantum number, 
contrary to its gluino $\tilde g'$ superpartner which has a negative
$R$-parity.}, it is
expected to decay mainly into light jets and/or top quarks, if kinematically
allowed. Let us note even if squarks and
gluinos could be lighter than the sgluon, which implies that additional
decay channels might become possible, this case is not considered in this work.
Subsequently, the superpartners and their interactions are irrelevant for the
phenomenological studies to be performed, and we therefore conceive a benchmark
scenario where they decouple.

In order to optimize the generation of the \madgraph\ model files, the masses of
the superpartners are set to a very high scale such as, \eg, 1000 TeV, and all
the mixing matrices related to sfermions, neutralinos and charginos are set to
zero. This makes all the interaction vertices involving two superpartners or
more to vanish, so that they are effectively removed from the generated UFO
model files by the optimization procedure described in Section
\ref{sec:UFOMG}. In addition to the Standard Model inputs, it is then enough to
fix the sgluon related parameters, \ie, its mass $m_\sigma$ and its couplings to
quarks and gluons $\lambda_L$, $\lambda_R$ and $\lambda_g$ defined in Eq.\
\eqref{eq:Leff}.
Let us note that even if the sgluon mass $m_\sigma$ depends, strictly
speaking, on several soft parameters, \ie, $m_2^G$, $m_3^G$ and $m_{\Phi_G}$,
we follow the simplifying approach to regroup all their contributions into a
single parameter $m_\sigma$.

To investigate sgluonic signatures at the LHC, we consider
a scenario where the sgluon is assumed to have a mass of 500 GeV, a rather
collider-friendly value leading to a largely visible pair-production cross
section at 7 TeV. The sgluon mass is thus high enough so that the $t \bar t$
decay channel is open (assuming non-vanishing $\lambda_L$ and/or $\lambda_R$
couplings). The production rate of a four top signature is hence
enhanced with respect to the Standard Model expectation of about 0.3 fb, and the
observation of such a signal might be a hint of the existence of a sgluon field. 
We impose that the sgluon singly couples only to top quarks and to gluons with
strengths given by $(\lambda_L)^3{}_3 = (\lambda_R)^3{}_3 = 0.3$ and 
$\lambda_g = 1.5\times 10^{-4}$, all other effective couplings being assumed
vanishing.

\begin{figure}[t]
\vspace{6cm}\hspace{-2cm}
\centerline{\psfig{file=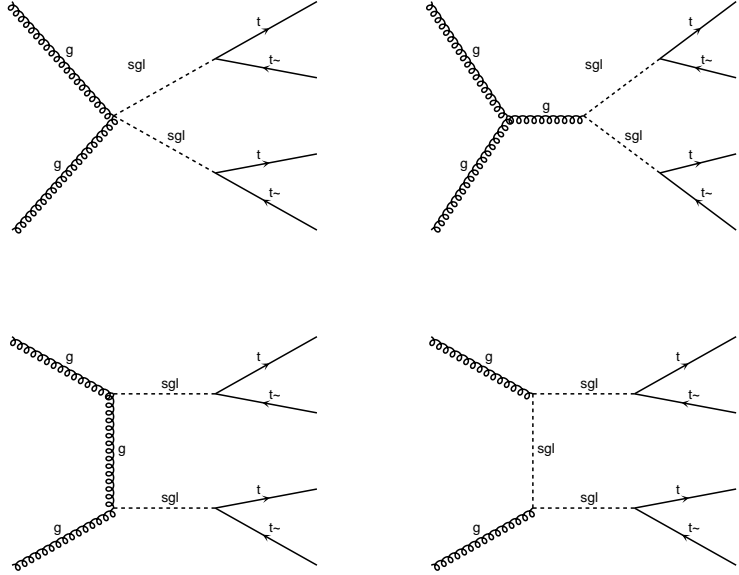,width=.75\columnwidth}}
\vspace*{8pt}
\vspace{-6.7cm}
\caption{Feynman diagrams associated to a four top signature issued from the
production of a pair of sgluons, as generated by \madgraph\ 5. In the notations above, the 
labels \texttt{t} and \texttt{t$\sim$} stand for the top quark and antiquark,
\texttt{g} for the gluon field and \texttt{sgl} for the sgluon. Let us note that
$u$-channel diagrams are omitted since they can be deduced from the
corresponding $t$-channel diagrams (lower panel). 
\protect\label{fig:sgluondiag}}
\end{figure}

The numerical values of the free parameters defining our simplified scenario are
summarized in Table \ref{tab:sglbenchmark}, together with the relevant Standard
Model inputs. In this scenario, the sgluon-pair production cross section reach
the level of 0.20 pb, whilst the sgluon-pair induced four top production cross
section is of about 42 fb, \ie, more than about 140 times the Standard Model
predictions. The corresponding Feynman diagrams, as generated by \madgraph\ 5,
are presented in Figure \ref{fig:sgluondiag}. Whereas the value of the effective
sgluon-gluon-gluon coupling of $1.5\times 10^{-4}$ is rather reduced, it
contributes significantly to sgluon-pair production and decays.
Indeed, the first Feynman diagram of the second line of 
Fig.\ \ref{fig:sgluondiag},
involving two effective vertices, contributes to the sgluon-pair total
production cross section by about 15\%. More importantly, the sgluon branching 
ratio to a gluon pair cannot be neglected, reaching more than 
50\%, due to the phase-space suppression of the top-antitop pair channel. 
 
\renewcommand{\arraystretch}{1.3}
\begin{table}[h]
  \tbl{Input parameters associated to the chosen benchmark scenario in
    the context of investigating sgluon production in the minimal $R$-symmetric
    supersymmetric model. We remind that the superpartners are decoupled and
    thus irrelevant and that all the other $\lambda_{\{L,R\}}$ parameters are
    vanishing.}
  {\begin{tabular}{|ccc|cccc|}
    \hline  $m_t$ [GeV] &$m_Z$ [GeV] & $\alpha_s(m_Z)$ & 
       $m_\sigma$ [GeV] & $(\lambda_L)^3{}_3$ &$(\lambda_R)^3{}_3$ & $\lambda_g$ \\
    \hline
    173.1 & 91.1876 & 0.1176 & 500 & 0.3 & 0.3 & 0.00015 \\ 
   \hline
   \end{tabular} \label{tab:sglbenchmark} }
\end{table}
\renewcommand{\arraystretch}{1.}

\section{Non-minimal supersymmetric phenomenology}\label{sec:pheno}

Event simulation is performed for the LHC collider at a center-of-mass energy of 
$\sqrt{s} = 7$ TeV and for an integrated luminosity of 4 fb$^{-1}$.
Concerning both signal and background events, hard scattering matrix elements
are described with the Monte Carlo generator \madgraph\ 5 \cite{Alwall:2011uj}.
Neglecting all quark masses but the top mass, we employ the leading order set of
the CTEQ6 parton density fit \cite{Pumplin:2002vw} and identify both the
renormalization and factorization scales as the transverse mass of the produced
heavy particles. 
We then match the events generated by \madgraph\ with parton showering
and hadronization as provided by the \pythia\ program. The version six of
\pythia\ \cite{Sjostrand:2006za} is used for background and sgluon signal
events, whilst the version 8 \cite{Sjostrand:2007gs} is used for $R$-parity
violating monotop signal events due to the exotic color structure not compliant
with the requirements of \pythia\ 6. We finally perform a fast detector
simulation with the program \delphes\ \cite{Ovyn:2009tx}, using the publicly
available CMS detector card. Jets are reconstructed using the anti-$k_{t}$
algorithm with a radius parameter of $R=0.5$. The two examples of
phenomenological analyses presented in this Section are performed with the help
of the \madanalysis\ 5 package \cite{ma5}.

\subsection{Monotop production in $R$-parity violating
supersymmetry} \label{sec:phenomonotop}
The main signatures associated with monotop production can be classified
according to the top quark decay,
\be
  p p \to t + \widetilde\chi^0_1 \to  b j j + \slashed{E}_T
    \quad \text{or} \quad b \ell + \slashed{E}_T \ ,
\ee
where $j$ and $b$ denote light and $b$-jets, respectively and $\ell$
a charged lepton. The missing transverse energy $\slashed{E}_T$ is associated to
the lightest neutralino $\widetilde \chi_1^0$ escaping the detector invisibly
due to its long lifetime \cite{Allanach:1999bf}, as well as to the neutrino
appearing in the case of leptonically decaying top quarks. Since
leptonically decaying monotop signatures induced by $R$-parity violating
supersymmetry have been widely investigated in the past 
\cite{Berger:1999zt,Berger:2000zk}, we therefore focus on monotop events where
the top quark decays hadronically.

The only source of irreducible Standard Model background to an hadronic
monotop signal consists in the production of an invisibly decaying $Z$-boson
together with at least three jets, one of them being tagged as a $b$ jet. 
In contrast, there are many possible sources of (dominant) instrumental
background related
to detector effects. On the one hand, QCD multijet events with misreconstructed
jets produce fake missing energy and mimic hence the monotop
signature. However, asking for the reconstruction of a top quark could help to
reject most of these QCD events. On the other hand, $W$ plus jets, $t\bar{t}$ and
diboson events where the weak bosons decay to nonreconstructed leptons, as well
as single top events including non-reconstructed or misrecontructed jets might
also be a possible source of background.

In recent experimental analyses \cite{daCosta:2011qk,Collaboration:2011ida}, it
has been shown that simple selection cuts allow to keep a good control on the
background. Inspired, in addition, by the parton-level results of Ref.\
\refcite{Andrea:2011ws}, we start by requiring a large amount of missing
transverse energy $\slashed{E}_T > 200$ GeV. Within the employed
simplified detector simulation, this
selection cut allows to remove all the QCD multijet events from the background
and to sensibly to reduce the contribution of the $t \bar t$ channel.
We then impose a veto on the presence in the final
state of any charged lepton (electron or muon) with a transverse momentum $p_T >
10$ GeV and a pseudorapidity $|\eta|<2.5$. This last selection cut does not 
affect the signal but sensibly reduce the contributions of $Z$-boson plus jets
and $W$-boson plus jets events.

In a second stage, we exploit the presence of a hadronically decaying top quark,
together with the one of its decay products. We hence demand exactly
one $b$-tagged jet with a transverse momentum $p_T > 50$ GeV and a
pseudorapidity $|\eta| < 2.5$, as well as exactly two light jets with a
transverse momentum $p_T > 30$ GeV and a pseudorapidity $|\eta| < 2.5$.
We estimate a $b$-tagging efficiency depending on the transverse momentum of the
jet as presented on Figure 3 (left panel) of Ref.\ \refcite{BTV-1}, together
with a charm and light jet mistagging rate (depending also on the four-momentum of
the jet) as on Figure 6 (right panel) of Ref.\ \refcite{BTV-2}. Consequently,
the efficiency of correctly tagging a jet with a transverse momentum of 50 GeV 
as a $b$-jet is of about 70\%, whilst the mistagging rate of a charm (light) jet
as a $b$-jet is of about 40 \% (2\%). In
addition, since the two light jets are issued from a $W$-boson, we constrain
their 
invariant-mass $m_{jj}$ to be compatible with the mass of the $W$-boson, \ie,
$m_{jj} \in [ m_W - 15$ GeV, $m_W + 15$ GeV $ ]$, with $m_W = 80.31$ GeV.

\begin{figure}[t]
\centerline{\psfig{file=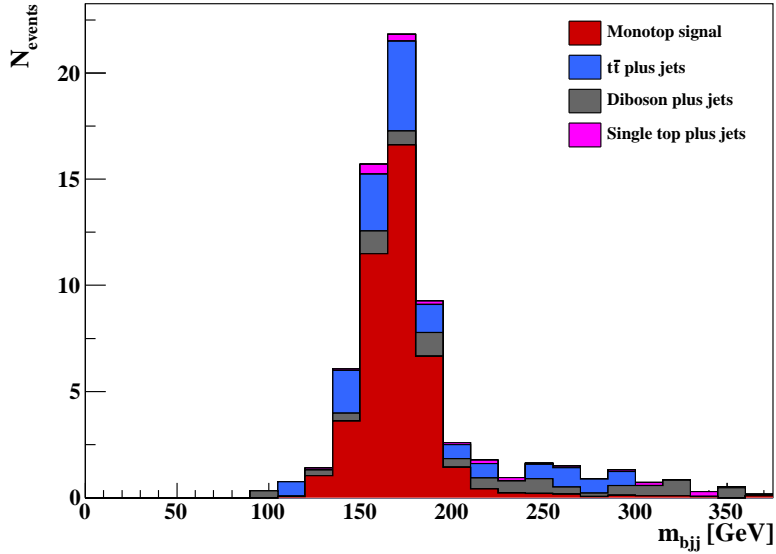,width=.9\columnwidth}}
\vspace*{8pt}
\caption{We demand a monotop configuration of the final state as described in
the text (large
missing energy, no charged leptons, exactly one $b$-tagged jet and two
light jets whose the invariant-mass is compatible with a $W$-boson). 
After applying those selection cuts, we present the invariant-mass
distribution 
of the three jets $m_{bjj}$ both for the signal (red) and the dominant sources
of background, \ie, single top (purple), $t \bar t$ (blue) and diboson (gray)
events.
\protect\label{fig:monotop}}
\end{figure}

The distribution of the invariant-mass $m_{bjj}$ of the three jets is
illustrated on Figure \ref{fig:monotop}. After applying all the cuts described
above, the dominant contributions to the Standard Model background are $t \bar t$
plus jets, diboson plus jets and single top plus jets events. In contrast, the
contributions of all the other sources of background, such as $W$-boson,
$Z$-boson or QCD events, are reduced to a barely visible level and thus under
a very good control. We further constrain  the system of the three selected jets
by requiring their invariant-mass $m_{bjj}$ to be compatible with the mass of the
top quark, lying thus in the range $[ m_t - 20$ GeV, $m_t + 20$ GeV $ ]$. The
final number of selected events is shown in Table
\ref{tab:monotop} both for the signal and all the sources of background,
together with the LHC sensitivity to a monotop signal. We define the latter 
as the number of signal events over the total number of
(signal plus background) events $S/\sqrt{S+B}$, which reaches in our scenario
the level of $4.95 \sigma$.

\renewcommand{\arraystretch}{1.3}
\begin{table}[h]
  \tbl{Number of remaining events $N_{\rm events}$ for both the different
    sources of background and for the signal after the whole set of selection
    cuts described in the text. Contributions of the $W$-boson plus jets, the
    $Z$-boson plus jets and the QCD multijet channels are zero and thus not
    indicated. The results correspond to an integrated luminosity of
    4 fb$^{-1}$ of proton-proton collisions at the LHC collider, running with a
    center-of-mass energy of 7 TeV. The LHC sensitivity, defined as the number of
    signal events over the total number of (signal plus background) events
    $S/\sqrt{S+B}$ is also indicated.}
  {\begin{tabular}{|c|c|}
    \hline  Event sample & $N_{\rm events}$ after all the selection cuts
      presented in the text\\
    \hline  \hline
    Top-antitop pairs plus jets & $8.2 \pm 2.3$ \\ 
    Diboson plus jets           & $2.7 \pm 0.7$ \\
    Single top                  & $0.9 \pm 0.3$ \\
   \hline
    Total background & $11.8 \pm 2.4$ \\
    Monotop signal   & $33.2 \pm 1.0$ \\
    Sensitivity     & $4.95$ \\
   \hline
   \end{tabular} \label{tab:monotop} }
\end{table}
\renewcommand{\arraystretch}{1.}

Conversely, a $3\sigma$ deviation from the Standard Model expectation could be 
already 
observed for any value of the $R$-parity violating parameter $\lpp_{312}
> 0.11$, assuming the supersymmetric spectrum to be unchanged. Since the
number of signal events is expected to be of the same order of magnitude
for moderate superpartner masses (below or around the TeV scale), the set of
selection cuts
presented above, \ie, standard monotop cuts, would be sufficient to render 
$R$-parity violating supersymmetric monotop signatures distinguishable for a
large region of the constrained MSSM parameter space.

\subsection{Multitop production in minimal $R$-symmetric supersymmetry}
The top-enriched signature described in Section \ref{sec:benchmrssm} leads to
final states with a large multiplicity of jets and leptons as the decay products
of the four top quarks. Therefore, the main sources of Standard Model
background is expected to be related to rare processes such as the production of
four top quarks or the production of a top-antitop pair in association with one
or several gauge bosons. 

In order
to reject a good fraction of the Standard Model background events, we require
the presence of exactly two charged leptons carrying the same
electric charge\footnote{In our simplified detector simulation, the charge of a
lepton is always correctly identified, which is far from being the 
case in experiments such as ATLAS or CMS.}, with 
a transverse momentum $p_T > 20$ GeV and a pseudorapidity $|\eta| < 2.5$. 
In addition, we require isolation criteria and reject leptons in the case they
are at a relative distance $\Delta R<0.2$ of a jet.

Moreover, each of the leptons is issued from a leptonically decaying top
quark. Therefore, it comes accompanied with a neutrino, carrying missing
transverse energy, and we then ask for a missing transverse energy selection cut,
keeping an event only if $\slashed{E}_T > 40$ GeV.

As stated above, our signal events are rather rich in jets. In particular, we
expect at least four $b$-tagged jets (one for each produced top quark) and four 
additional light jets issue from the hadronically decaying top quarks.
Consequently, we only select events with at least eight jets, each of them
having a transverse energy $E_{T}>20$ GeV, and at least three of the jets are
required to be $b$-tagged. As in Section \ref{sec:phenomonotop}, our
$b$-tagging efficiency is estimated as on Figure 3 (left panel) of Ref.\
\refcite{BTV-1}, together with a charm and light jet mistagging rate as on
Figure 6 (right panel) of Ref.\ \refcite{BTV-2}. We remind that all the
efficiencies depend on the transverse momentum of the jet. 

\begin{figure}[t]
\centerline{\psfig{file=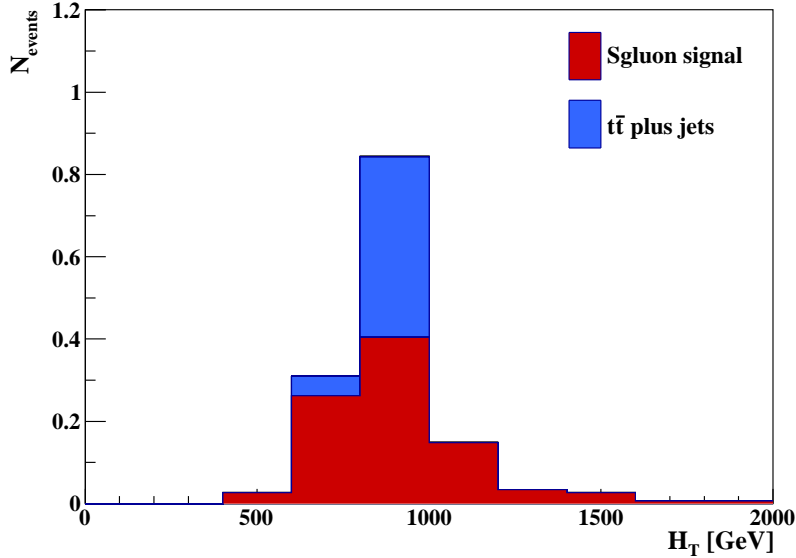,width=.9\columnwidth}}
\vspace*{8pt}
\caption{Applying selection cuts related to a final state with an important
hadronic activity (at least eight jets, three $b$-tagged jets) and containing two
leptons with the same electric charge and a sensible missing energy, we present
the distribution of the $H_T$ variable defined as the scalar sum of the
transverse momentum of all the selected leptons and jets both for the signal
(red) and the dominant source of background, \ie, $t \bar t$ (blue) events.
\protect\label{fig:4top}}
\end{figure}

The important hadronic activity in the final state suggests to
investigate the $H_T$ variable, defined as the scalar sum of the
transverse momentum of all the selected leptons and jets. The results are
presented in Figure \ref{fig:4top}. After applying the set of cuts described
above, the dominant contributions to the background consist in $t\bar t$ events,
as well as, in a smaller (and negligible) extent, 
in $t \bar t$ plus one or several gauge bosons
events. This is also presented in Table \ref{tab:sgluon} where we have omitted
the non-contributing sources of background. 

\begin{figure}[t]
  \centerline{\psfig{file=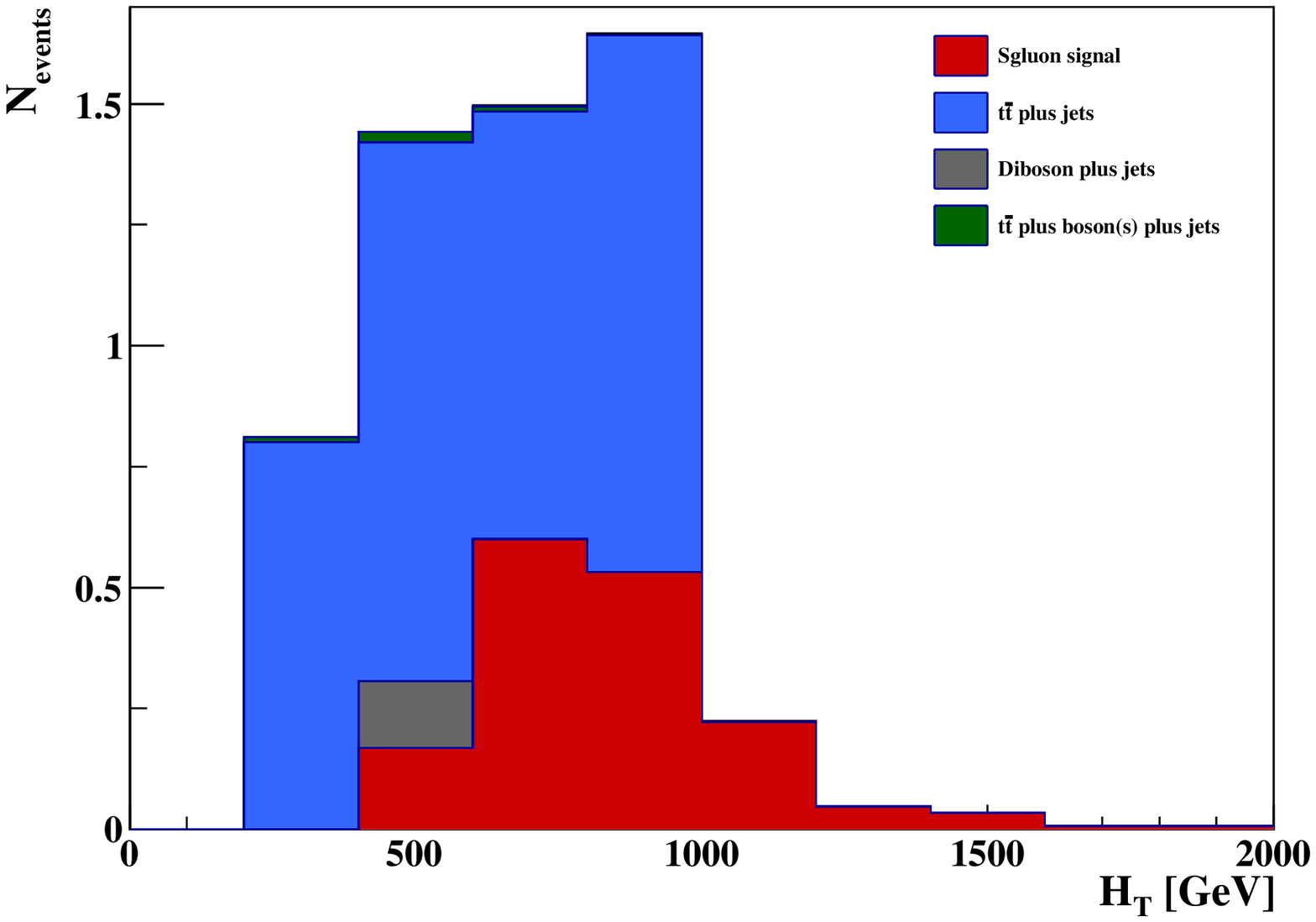,width=.9\columnwidth}}
  \centerline{\psfig{file=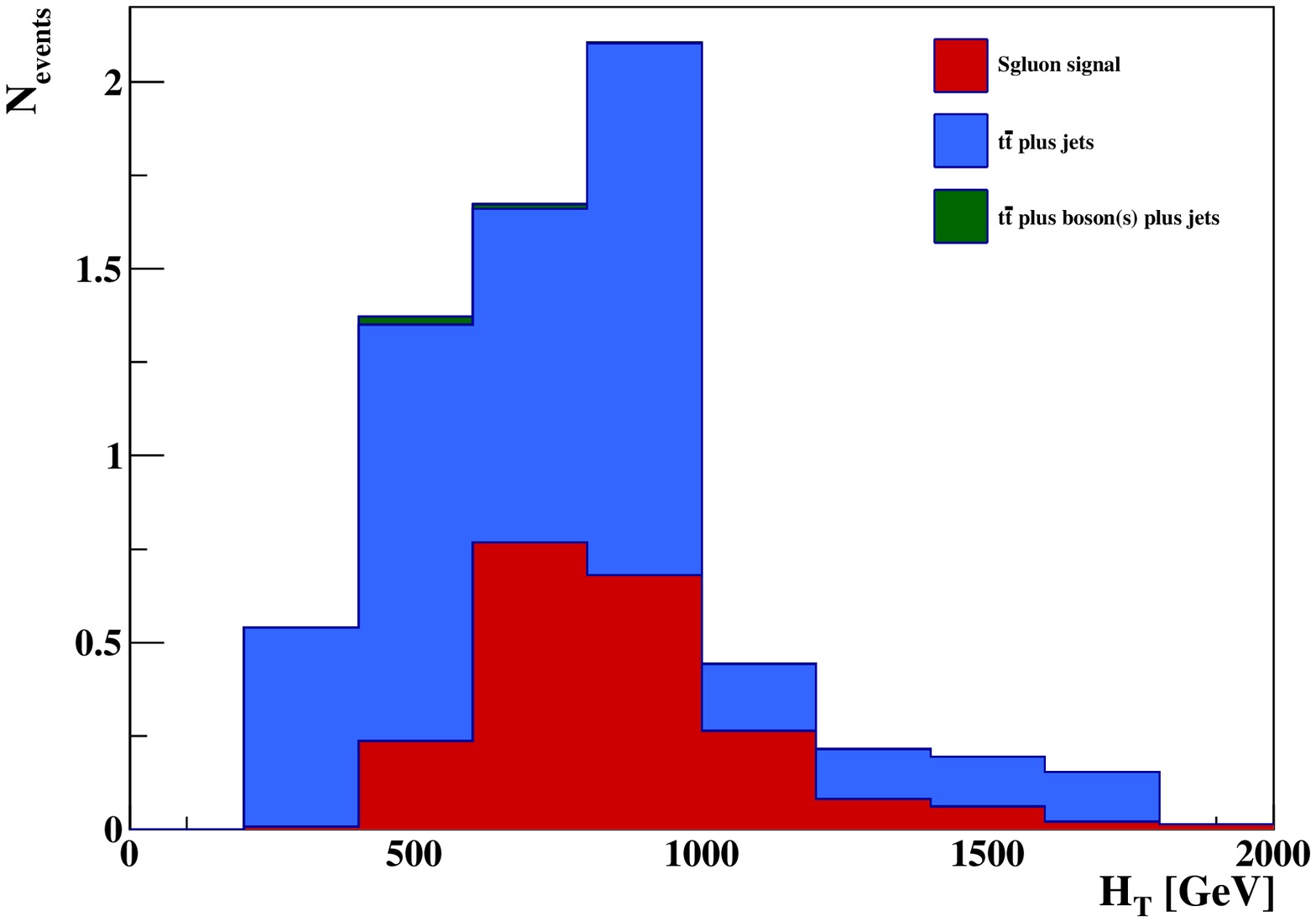,width=.9\columnwidth}}
\vspace*{8pt}
\caption{Same as Figure \ref{fig:4top}, but with a weaker requirement of only
six (seven) jets, three (two) of them being $b$-tagged jets in the upper (lower)
panel of the figure. Events related to the production of a $t \bar t$ pair in
association with one or several additional gauge boson are shown in green,
whilst diboson events are shown in gray.
\protect\label{fig:4topb}}
\end{figure}

Our event selection
criteria are very restrictive. However, due to the smallness of the signal cross
section, these cuts are mandatory to ensure a good background rejection. In the
case of higher luminosity or more favorable benchmark scenarios, one could
however lower the requirements by demanding less jets with a transverse
energy $E_{T}>20$, and/or a smaller number of $b$-tagged jets. The example of
asking for six (seven) jets including three (two) $b$-tagged jets is shown in
the upper (lower) panel of Figure \ref{fig:4topb}. In this case, the background
clearly dominates over the signal.

\renewcommand{\arraystretch}{1.3}
\begin{table}[h]
  \tbl{Number of events $N_{\rm events}$ for both the different
    sources of background and for the signal after the whole set of selection
    cuts described in the text. Contributions of the single and diboson plus
    jets, the associated production of a $t \bar t$ pair and one or several
    gauge bosons,the single top plus jets and the QCD multijet channels are zero
    and thus not indicated.  The results 
    correspond to an integrated luminosity of
    4 fb$^{-1}$ of proton-proton collisions at the LHC collider, running with a
    center-of-mass energy of 7 TeV.}
  {\begin{tabular}{|c|c|}
    \hline  Event sample & $N_{\rm events}$ after the selection cuts presented
in the text\\
    \hline  \hline
    Top-antitop pairs plus jets & $0.5 \pm 0.3$ \\ 
   \hline
    Total background & $ 0.5 \pm 0.3$ \\
    Sgluon signal   & $0.9  \pm 0.1$ \\
   \hline
   \end{tabular} \label{tab:sgluon} }
\end{table}
\renewcommand{\arraystretch}{1.}

Even if the background rejection is pretty efficient with those cuts,
a 4 fb$^{-1}$ luminosity is not enough to obtain a sensitivity to the new
physics signal. Therefore, only lighter sgluons would be accessible at 7 TeV.
The results are however promising for the 2012 LHC run at a center-of-mass
energy of 8 TeV and with an expected luminosity of 15 fb$^{-1}$.

\section{Conclusions}\label{sec:conclusions}
In this work, we have illustrated how softly broken supersymmetric quantum field
theories can be efficiently and easily implemented in the \feynrules\ package,
exported then to Monte Carlo event generators and finally be ready for
phenomenological investigations at hadron colliders.

Thanks to its superspace module, implementing a supersymmetric theory in
\feynrules\ is reduced to the very simple task of providing the superfield and
field content of the model, defining the free parameters appearing in the
Lagrangian as well as providing the 
superpotential, in terms of superfields, and the soft supersymmetry breaking
Lagrangian. The \feynrules\ program allows then to automatically expand the
superfield Lagrangian in terms of the physical component fields, rendering it
ready to be exported either to Monte Carlo event generators or to a \python\
library dubbed the UFO model format, containing all the model information.

The UFO model format is agnostic of any assumption on the new physics model
under consideration. Any interaction vertex, whatever is the number of
incoming/outgoing particles, the included color and Lorentz structures, can be
effectively exported, in contrast to the other model format used by  
automated Monte Carlo tools. The latter indeed rely on text files to be parsed
and a set of hard-coded compliant color and Lorentz structures.

In this work, we have presented two phenomenological studies of collider
signatures associated to interaction vertices with exotic color structures.
Consequently, only the
\madgraph\ 5 Monte Carlo generator, using the UFO format, is capable of
generating events associated to such signatures. We therefore employ it to
investigate monotop production in the framework of the minimal supersymmetric
standard model with $R$-parity violation as well as multitop production in the
context of the minimal $R$-symmetric supersymmetric model. We show that for the 
chosen benchmark, the LHC is the perfect machine to unveil the presence of the
corresponding new physics.

\section*{Acknowledgements}
The author is extremely grateful to J.\ Andrea and E.\ Conte for their precious
support in the phenomenological analyses performed in this work and to O.\
Mattelaer for his help with the generation of the events. The author
also thanks the whole \feynrules, UFO and \madgraph\ 5 development teams, as
well as
J.\ Alwall, C.\ Duhr, F.\ Maltoni, K.\ Mawatari and  M.\ Rausch de
Traubenberg for discussions about the present manuscript.  This work was
supported by the Theory-LHC France Initiative of the CNRS/IN2P3.


\begin{thebibliography}{0}
\bibitem{Nilles:1983ge}
  H.~P.~Nilles,
  Phys.\ Rept.\  {\bf 110}, 1 (1984).

\bibitem{Haber:1984rc}
  H.~E.~Haber and G.~L.~Kane,
  Phys.\ Rept.\  {\bf 117}, 75 (1985).
\bibitem{ATLAS}
  https://twiki.cern.ch/twiki/bin/view/AtlasPublic/SupersymmetryPublicResults
\bibitem{CMS}
  https://twiki.cern.ch/twiki/bin/view/CMSPublic/PhysicsResultsSUS

\bibitem{Mangano:2002ea} 
  M.~L.~Mangano, M.~Moretti, F.~Piccinini, R.~Pittau and A.~D.~Polosa,
  JHEP {\bf 0307}, 001 (2003).

\bibitem{Gleisberg:2008fv} 
  T.~Gleisberg and S.~Hoeche,
  JHEP {\bf 0812}, 039 (2008).

\bibitem{Pukhov:1999gg}
  A.~Pukhov {\it et al.},
  arXiv:hep-ph/9908288.

\bibitem{Boos:2004kh}
  E.~Boos {\it et al.}  [CompHEP Collaboration],
  Nucl.\ Instrum.\ Meth.\  A {\bf 534}, 250 (2004).

\bibitem{Pukhov:2004ca}
  A.~Pukhov,
  arXiv:hep-ph/0412191.

\bibitem{Cafarella:2007pc} 
  A.~Cafarella, C.~G.~Papadopoulos and M.~Worek,
  Comput.\ Phys.\ Commun.\  {\bf 180}, 1941 (2009).

\bibitem{Stelzer:1994ta}
  T.~Stelzer and W.~F.~Long,
  Comput.\ Phys.\ Commun.\  {\bf 81}, 357 (1994).

\bibitem{Maltoni:2002qb}
  F.~Maltoni and T.~Stelzer,
  JHEP {\bf 0302}, 027 (2003).

\bibitem{Alwall:2007st}
  J.~Alwall {\it et al.},
  JHEP {\bf 0709}, 028 (2007).

\bibitem{Alwall:2008pm}
  J.~Alwall, P.~Artoisenet, S.~de Visscher, C.~Duhr, R.~Frederix, M.~Herquet and O.~Mattelaer,
  AIP Conf.\ Proc.\  {\bf 1078}, 84 (2009).

\bibitem{Alwall:2011uj} 
  J.~Alwall, M.~Herquet, F.~Maltoni, O.~Mattelaer and T.~Stelzer,
  JHEP {\bf 1106}, 128 (2011).

\bibitem{Gleisberg:2003xi} 
  T.~Gleisberg, S.~Hoeche, F.~Krauss, A.~Schalicke, S.~Schumann and J.~-C.~Winter,
  JHEP {\bf 0402}, 056 (2004).

\bibitem{Gleisberg:2008ta}
  T.~Gleisberg, S.~Hoche, F.~Krauss, M.~Schonherr, S.~Schumann, F.~Siegert and J.~Winter,
  JHEP {\bf 0902}, 007 (2009).

\bibitem{Moretti:2001zz}
  M.~Moretti, T.~Ohl and J.~Reuter,
  In *2nd ECFA/DESY Study 1998-2001* 1981-2009.

\bibitem{Kilian:2007gr} 
  W.~Kilian, T.~Ohl and J.~Reuter,
  Eur.\ Phys.\ J.\ C {\bf 71}, 1742 (2011).

\bibitem{Semenov:1998eb} 
  A.~Semenov,
  Comput.\ Phys.\ Commun.\  {\bf 115}, 124 (1998).

\bibitem{Semenov:2008jy} 
  A.~Semenov,
  Comput.\ Phys.\ Commun.\  {\bf 180}, 431 (2009).

\bibitem{Christensen:2008py} 
  N.~D.~Christensen and C.~Duhr,
  Comput.\ Phys.\ Commun.\  {\bf 180}, 1614 (2009).

\bibitem{Christensen:2009jx}
  N.~D.~Christensen, P.~de Aquino, C.~Degrande, C.~Duhr, B.~Fuks, M.~Herquet, F.~Maltoni and S.~Schumann,
  Eur.\ Phys.\ J.\ C {\bf 71} (2011) 1541.

\bibitem{Christensen:2010wz} 
  N.~D.~Christensen, C.~Duhr, B.~Fuks, J.~Reuter and C.~Speckner,
  arXiv:1010.3251 [hep-ph].

\bibitem{Duhr:2011se} 
  C.~Duhr and B.~Fuks,
  Comput.\ Phys.\ Commun.\  {\bf 182}, 2404 (2011).

\bibitem{Staub:2009bi} 
  F.~Staub,
  Comput.\ Phys.\ Commun.\  {\bf 181}, 1077 (2010)

\bibitem{Staub:2010jh} 
  F.~Staub,
  Comput.\ Phys.\ Commun.\  {\bf 182}, 808 (2011).

\bibitem{Salam:1974yz} 
  A.~Salam and J.~A.~Strathdee,
  Nucl.\ Phys.\ B {\bf 76}, 477 (1974).

\bibitem{Ferrara:1974ac} 
  S.~Ferrara, J.~Wess and B.~Zumino,
  Phys.\ Lett.\ B {\bf 51}, 239 (1974).

\bibitem{Hahn:1998yk} 
  T.~Hahn and M.~Perez-Victoria,
  Comput.\ Phys.\ Commun.\  {\bf 118}, 153 (1999).

\bibitem{Hahn:2000kx}
  T.~Hahn,
  Comput.\ Phys.\ Commun.\  {\bf 140} (2001) 418.

\bibitem{Degrande:2011ua} 
  C.~Degrande, C.~Duhr, B.~Fuks, D.~Grellscheid, O.~Mattelaer and T.~Reiter,
  arXiv:1108.2040 [hep-ph].

\bibitem{Cullen:2011ac} 
  G.~Cullen, N.~Greiner, G.~Heinrich, G.~Luisoni, P.~Mastrolia, G.~Ossola, T.~Reiter and F.~Tramontano,
  arXiv:1111.2034 [hep-ph].

\bibitem{deAquino:2011ub} 
  P.~de Aquino, W.~Link, F.~Maltoni, O.~Mattelaer and T.~Stelzer,
  arXiv:1108.2041 [hep-ph].

\bibitem{Murayama:1992gi} 
  H.~Murayama, I.~Watanabe and K.~Hagiwara,
  KEK-91-11.

\bibitem{Hagiwara:2008jb} 
  K.~Hagiwara, J.~Kanzaki, Q.~Li and K.~Mawatari,
  Eur.\ Phys.\ J.\ C {\bf 56}, 435 (2008).

\bibitem{Hagiwara:2010pi} 
  K.~Hagiwara, K.~Mawatari and Y.~Takaesu,
  Eur.\ Phys.\ J.\ C {\bf 71}, 1529 (2011).

\bibitem{Mawatari:2011jy} 
  K.~Mawatari and Y.~Takaesu,
  Eur.\ Phys.\ J.\ C {\bf 71}, 1640 (2011).


\bibitem{Barbier:2004ez} 
  R.~Barbier, C.~Berat, M.~Besancon, M.~Chemtob, A.~Deandrea, E.~Dudas, P.~Fayet and S.~Lavignac {\it et al.},
  Phys.\ Rept.\  {\bf 420}, 1 (2005).

\bibitem{Fayet:1975yi} 
  P.~Fayet,
  Nucl.\ Phys.\ B {\bf 113}, 135 (1976).

\bibitem{AlvarezGaume:1996mv} 
  L.~Alvarez-Gaume and S.~F.~Hassan,
  Fortsch.\ Phys.\  {\bf 45}, 159 (1997).

\bibitem{Choi:2008pi} 
  S.~Y.~Choi, M.~Drees, A.~Freitas and P.~M.~Zerwas,
  Phys.\ Rev.\ D {\bf 78}, 095007 (2008).

\bibitem{Choi:2008ub} 
  S.~Y.~Choi, M.~Drees, J.~Kalinowski, J.~M.~Kim, E.~Popenda and P.~M.~Zerwas,
  Phys.\ Lett.\ B {\bf 672}, 246 (2009).

\bibitem{Choi:2009jc} 
  S.~Y.~Choi, M.~Drees, J.~Kalinowski, J.~M.~Kim, E.~Popenda and P.~M.~Zerwas,
  Acta Phys.\ Polon.\ B {\bf 40}, 1947 (2009).

\bibitem{Choi:2010gc} 
  S.~Y.~Choi, D.~Choudhury, A.~Freitas, J.~Kalinowski, J.~M.~Kim and P.~M.~Zerwas,
  JHEP {\bf 1008}, 025 (2010).

\bibitem{Schumann:2011ji} 
  S.~Schumann, A.~Renaud and D.~Zerwas,
  JHEP {\bf 1109}, 074 (2011).

\bibitem{Salam:1974xa} 
  A.~Salam and J.~A.~Strathdee,
  Nucl.\ Phys.\ B {\bf 87}, 85 (1975).

\bibitem{Fayet:1974pd} 
  P.~Fayet,
  Nucl.\ Phys.\ B {\bf 90}, 104 (1975).

\bibitem{Kribs:2007ac} 
  G.~D.~Kribs, E.~Poppitz and N.~Weiner,
  Phys.\ Rev.\ D {\bf 78}, 055010 (2008).

\bibitem{Choi:2010an}
  S.~Y.~Choi, D.~Choudhury, A.~Freitas, J.~Kalinowski and P.~M.~Zerwas,
  Phys.\ Lett.\ B {\bf 697}, 215 (2011)
  [Erratum-ibid.\ B {\bf 698}, 457 (2011)].

\bibitem{Plehn:2008ae} 
  T.~Plehn and T.~M.~P.~Tait,
  J.\ Phys.\ G G {\bf 36}, 075001 (2009).

\bibitem{Brignole:2003cm} 
  A.~Brignole, J.~A.~Casas, J.~R.~Espinosa and I.~Navarro,
  Nucl.\ Phys.\ B {\bf 666}, 105 (2003).

\bibitem{Carena:2009gx} 
  M.~Carena, K.~Kong, E.~Ponton and J.~Zurita,
  Phys.\ Rev.\ D {\bf 81}, 015001 (2010).

\bibitem{Antoniadis:2009rn} 
  I.~Antoniadis, E.~Dudas, D.~M.~Ghilencea and P.~Tziveloglou,
  Nucl.\ Phys.\ B {\bf 831}, 133 (2010).

\bibitem{FuksRausch}
  B.~Fuks and M.~Rausch de Traubenberg, \textit{Supersym\'etrie~: exercices avec
    solutions} (Ed.\ Ellipses, 2011).


\bibitem{Cremmer:1978hn} 
  E.~Cremmer, B.~Julia, J.~Scherk, S.~Ferrara, L.~Girardello and P.~van Nieuwenhuizen,
  Nucl.\ Phys.\ B {\bf 147}, 105 (1979).

\bibitem{Zumino:1979et} 
  B.~Zumino,
  Phys.\ Lett.\ B {\bf 87}, 203 (1979).

\bibitem{Cremmer:1982wb} 
  E.~Cremmer, S.~Ferrara, L.~Girardello and A.~Van Proeyen,
  Phys.\ Lett.\ B {\bf 116}, 231 (1982).

\bibitem{Cremmer:1982en} 
  E.~Cremmer, S.~Ferrara, L.~Girardello and A.~Van Proeyen,
  Nucl.\ Phys.\ B {\bf 212}, 413 (1983).

\bibitem{'tHooft:1976up} 
  G.~'t Hooft,
  Phys.\ Rev.\ Lett.\  {\bf 37}, 8 (1976).

\bibitem{Dimopoulos:1988jw}
  S.~Dimopoulos and L.~J.~Hall,
  Phys.\ Lett.\  B {\bf 207}, 210 (1988).

\bibitem{Barger:1989rk}
  V.~D.~Barger, G.~F.~Giudice and T.~Han,
  Phys.\ Rev.\  D {\bf 40}, 2987 (1989).

\bibitem{Intriligator:2006dd} 
  K.~A.~Intriligator, N.~Seiberg and D.~Shih,
  JHEP {\bf 0604}, 021 (2006).

\bibitem{Intriligator:2007py} 
  K.~A.~Intriligator, N.~Seiberg and D.~Shih,
  JHEP {\bf 0707}, 017 (2007).

\bibitem{Polchinski:1982an} 
  J.~Polchinski and L.~Susskind,
  Phys.\ Rev.\ D {\bf 26}, 3661 (1982).

\bibitem{Dine:1992yw} 
  M.~Dine and D.~MacIntire,
  .Phys.\ Rev.\ D {\bf 46}, 2594 (1992).

\bibitem{Fox:2002bu} 
  P.~J.~Fox, A.~E.~Nelson and N.~Weiner,
  JHEP {\bf 0208}, 035 (2002).

\bibitem{Corcella:2000bw} 
  G.~Corcella, I.~G.~Knowles, G.~Marchesini, S.~Moretti, K.~Odagiri, P.~Richardson, M.~H.~Seymour and B.~R.~Webber,
  JHEP {\bf 0101}, 010 (2001).

\bibitem{Bahr:2008pv} 
  M.~Bahr, S.~Gieseke, M.~A.~Gigg, D.~Grellscheid, K.~Hamilton, O.~Latunde-Dada, S.~Platzer and P.~Richardson {\it et al.},
  Eur.\ Phys.\ J.\ C {\bf 58}, 639 (2008).

\bibitem{Sjostrand:2006za} 
  T.~Sjostrand, S.~Mrenna and P.~Z.~Skands,
  JHEP {\bf 0605}, 026 (2006).

\bibitem{Sjostrand:2007gs}
  T.~Sjostrand, S.~Mrenna and P.~Z.~Skands,
  Comput.\ Phys.\ Commun.\  {\bf 178}, 852 (2008).

\bibitem{AbdusSalam:2011fc} 
  S.~S.~AbdusSalam, B.~C.~Allanach, H.~K.~Dreiner, J.~Ellis, U.~Ellwanger, J.~Gunion, S.~Heinemeyer and M.~Kraemer {\it et al.},
  Eur.\ Phys.\ J.\ C {\bf 71}, 1835 (2011).

\bibitem{Lancaster:2011wr} 
  M.~Lancaster [Tevatron Electroweak Working Group and for the CDF and D0 Collaborations],
  arXiv:1107.5255 [hep-ex].

\bibitem{Nakamura:2010zzi} 
  K.~Nakamura {\it et al.}  [Particle Data Group Collaboration],
  J.\ Phys.\ G G {\bf 37}, 075021 (2010).


\bibitem{Barbieri:1985ty}
  R.~Barbieri and A.~Masiero,
  Nucl.\ Phys.\  B {\bf 267}, 679 (1986).

\bibitem{Abel:1996qj}
  S.~A.~Abel,
  Phys.\ Lett.\  B {\bf 410}, 173 (1997).

\bibitem{Slavich:2000xm}
  P.~Slavich,
  Nucl.\ Phys.\  B {\bf 595}, 33 (2001).

\bibitem{Chakraverty:2000df}
  D.~Chakraverty and D.~Choudhury,
  Phys.\ Rev.\  D {\bf 63}, 112002 (2001).

\bibitem{BarShalom:2002sv}
  S.~Bar-Shalom, G.~Eilam and Y.~D.~Yang,
  Phys.\ Rev.\  D {\bf 67}, 014007 (2003).

\bibitem{Zwirner:1984is}
  F.~Zwirner,
  Phys.\ Lett.\  B {\bf 132}, 103 (1983).

\bibitem{Dimopoulos:1987rk}
  S.~Dimopoulos and L.~J.~Hall,
  Phys.\ Lett.\  B {\bf 196}, 135 (1987).

\bibitem{Hinchliffe:1992ad}
  I.~Hinchliffe and T.~Kaeding,
  Phys.\ Rev.\  D {\bf 47}, 279 (1993).

\bibitem{Chang:1996sw}
  D.~Chang and W.~Y.~Keung,
  Phys.\ Lett.\  B {\bf 389}, 294 (1996).

\bibitem{Choi:1998ak}
  K.~Choi, K.~Hwang and J.~S.~Lee,
  Phys.\ Lett.\  B {\bf 428} 129, (1998).

\bibitem{Bhattacharyya:1998dt}
  G.~Bhattacharyya and P.~B.~Pal,
  Phys.\ Lett.\  B {\bf 439}, 81 (1998).

\bibitem{Baltz:1997ar}
  E.~A.~Baltz and P.~Gondolo,
  Phys.\ Rev.\  D {\bf 57}, 7601 (1998).

\bibitem{Porod:2003um}
  W.~Porod,
  Comput.\ Phys.\ Commun.\  {\bf 153}, 275 (2003).

\bibitem{Porod:2011nf} 
  W.~Porod and F.~Staub,
  arXiv:1104.1573 [hep-ph].

\bibitem{Pumplin:2002vw} 
  J.~Pumplin, D.~R.~Stump, J.~Huston, H.~L.~Lai, P.~M.~Nadolsky and W.~K.~Tung,
  JHEP {\bf 0207}, 012 (2002).

\bibitem{Allanach:1999bf} 
  B.~Allanach {\it et al.}  [R parity Working Group Collaboration],
  hep-ph/9906224.

\bibitem{Andrea:2011ws} 
  J.~Andrea, B.~Fuks and F.~Maltoni,
  Phys.\ Rev.\ D {\bf 84}, 074025 (2011).

\bibitem{Ovyn:2009tx} 
  S.~Ovyn, X.~Rouby and V.~Lemaitre,
  arXiv:0903.2225 [hep-ph].

\bibitem{ma5} 
  E.~Conte, B.~ Fuks and G.~Serret,
  in preparation.

\bibitem{Berger:1999zt} 
  E.~L.~Berger, B.~W.~Harris and Z.~Sullivan,
  Phys.\ Rev.\ Lett.\  {\bf 83}, 4472 (1999).

\bibitem{Berger:2000zk} 
  E.~L.~Berger, B.~W.~Harris and Z.~Sullivan,
  Phys.\ Rev.\ D {\bf 63}, 115001 (2001).

\bibitem{daCosta:2011qk} 
  G.~Aad {\it et al.}  [Atlas Collaboration],
  Phys.\ Lett.\ B {\bf 701}, 186 (2011).

\bibitem{Collaboration:2011ida} 
  S.~Chatrchyan {\it et al.}  [CMS Collaboration],
  JHEP {\bf 1108}, 155 (2011).

\bibitem{BTV-1}
  CMS Collaboration, CMS-BTV-11-001.

\bibitem{BTV-2}
  CMS Collaboration, CMS-BTV-09-001.


\end{thebibliography}
\end{document}